\documentclass[preprint]{aastex}

\usepackage{amsmath,amssymb,graphicx}

\shorttitle{Astrophysical Fluids via DSS}
\shortauthors{Tobias et al.}

\begin{document}


\title{Astrophysical Fluid Dynamics via Direct Statistical Simulation}


\author{S.M. Tobias}
\affil{Department of Applied Mathematics, University of Leeds, Leeds,
  LS2 9JT, U.K.}
\email{smt@maths.leeds.ac.uk}

\and

\author{K. Dagon and J.B. Marston}
\affil{Department of Physics, Brown University, Providence, Rhode Island, 02912-1843 USA.}
%\email{marston@brown.edu}


%\altaffiltext{1}{Visiting Astronomer, Cerro Tololo Inter-American Observatory.
%CTIO is operated by AURA, Inc.\ under contract to the National Science
%\altaffiltext{2}{Society of Fellows, Harvard University.}
%\altaffiltext{3}{present address: Center for Astrophysics,
%    60 Garden Street, Cambridge, MA 02138}
%\altaffiltext{4}{Visiting Programmer, Space Telescope Science Institute}
%\altaffiltext{5}{Patron, Alonso's Bar and Grill}


\begin{abstract}
In this paper we introduce the concept of Direct Statistical Simulation (DSS) for astrophysical flows. 
This technique may be appropriate for problems in astrophysical fluids where the instantaneous 
dynamics of the flows are of secondary importance to their statistical properties. We give examples of such problems including mixing and transport in planets, stars and disks. The method is described for a general set of evolution equations, before we consider the specific case of a spectral method optimised for problems on a spherical surface. The method is illustrated for the simplest non-trivial example of hydrodynamics and MHD on a rotating spherical surface. We then discuss possible extensions of the method both in terms of computational methods and the range of astrophysical problems that are of interest.
\end{abstract}

\section{Introduction}

\label{intro}
%\subsection{}
%\subsubsection{}

The modeling of astrophysical phenomena is often limited by the huge
range of spatial and temporal scales that need to be resolved in order
to describe accurately the dynamics. In many cases the large-scale behavior of
cosmic bodies depends on interactions at smaller scales that need to
be represented properly for a complete understanding of the
astrophysical phenomenon in question. The situation is usually
complicated by the requirement of including the back-reaction of the
large-scale environment on the smaller scale dynamics in a
self-consistent manner.
These types of problems are ubiquitous in astrophysics, but we list
here some important examples. The transport of angular momentum in
accretion disks may be mediated by the (magneto-)rotational turbulence that
is present in the disk \citep[see e.g.][]{balhaw:1998}. This
turbulence is itself driven and modulated by the large-scale
environment of Keplerian rotation and large-scale magnetic fields \citep[see e.g.][]{jamjulknob:2008}. The differential rotation pattern in stars (including the
sun) arises through an interaction of buoyancy-driven turbulence and
rotation, with Reynolds stresses at intermediate scales leading to
correlations that drive large-scale flows that themselves act back on
the turbulence \citep{ruediger:1989,brunetal:2002,rempel:2005,miesch:2005}. This
situation is mirrored in planets, where convective processes may create stresses
leading to large-scale flows.  Such stresses create turbulence in stably
stratified outer weather layers (for example  in Jupiter and Saturn) that may drive
the formation of jets \citep[see e.g.][and the references therein]{scottpolvani:2008}.

There are a number of approaches to modeling the fluid interactions
in astrophysical objects.  The approach taken often depends on whether it is the dynamics or statistics of the system that is of interest. Sometimes information about the dynamics --- that is the precise evolution of a particular realisation of a system --- is sometimes required for prediction or to compare with observations. More likely is that the statistics --- i.e.\ the average properties of an ensemble of evolutions --- is of interest; this may give more insight into the underlying physics of the system. 

Theoretically and computationally, a natural first approach is to perform direct numerical
simulations (DNS) of the fluid (or MHD) equations for the system. This approach is
the most straight-forward and has led to breakthroughs in many
branches of astrophysical fluid dynamics. This approach lends itself naturally to determining the dynamics of a given system.  However the extreme nature of the astrophysical 
turbulent environment ensures that not all spatial scales may be
faithfully represented even on the most massive parallel computers
available today. For this reason the practitioners of DNS must accept
that they are not in the correct parameter regime or  may claim that
the parameters take into account the effects of scales below the grid
cut-off via eddy diffusivities (sometimes termed turbulent transport
coefficients). These diffusivities are usually chosen in a plausible
but ad-hoc manner. Moreover, DNS may not be an efficient algorithm for determining statistics, since the ensemble over a large number of expensive calculations may be required in order to achieve meaningful statistics.

An alternative approach, which is not useful for determining dynamics but may be useful for statistics, is to derive evolution equations for the
large-scale dynamics and to formulate closure models for net effects
of the dynamics at moderate and small scales. Such models have a long
history in astrophysics and have also achieved some
measure of success (see e.g. \cite{kitchrued:1995,ogilvie:2003,rempel:2005}).  
This approach often utilises (either implicitly or explicitly) moment hierarchies (see e.g. \cite{Canuto:1994p890,Canuto:2001p899,farrell:2008p3353,Garaud:2010p911}). In particular it is customary to relate the average of local interactions of the small scale to the local values of the large-scale fields.
The weakness of such models is that they usually
rely on some ad-hoc assumption to close the model --- parameterising the interactions between large and small scales --- and often make the
assumption of homogeneity
or isotropy. They sometimes find it difficult to include
self-consistently the effects of the dynamic large-scale
environment. Often it is the case in astrophysics that regions of
strong transport lie in close proximity to regions of weak or no
transport or mixing --- for a nice example see the jets in Jupiter ---
and so closures that rely on homogeneity may lead to misleading
large-scale dynamics. Moreover it is often the case that the inclusion of such
closure models introduces new adjustable parameters to the problem
that can be tuned to fit observations and that little is known about
the sensitivity of the large-scale dynamics to changes in the
parameterizations of the scales not captured.

In this paper, we present a new approach to the problem of describing
astrophysical flows with a range of spatial scales, which we believe
will prove useful for a certain class of problems with large-scale
inhomogeneous and anisotropic flows. Specifically we describe the
development of efficient numerical algorithms to solve truncated
hierarchies of cumulant equations, leading directly to the statistical 
description of astrophysical flows, and we show that this direct statistical 
simulation (DSS) is able to reproduce qualitatively statistics obtained by 
time averaging DNS\footnote{We note here that we choose the terminology direct statistical simulation as we are solving for the statistics directly.}. That DSS has several advantages over DNS has
long been recognized, going back at least as far as a seminal monograph by 
\cite{Lorenz67}.   Low-order statistics are smoother in space and stiffer in
time than the underlying detailed flows.   Statistically stationary fixed points or slowly varying statistics can 
therefore be described with fewer degrees of freedom and also can be accessed more rapidly.
Convergence with increasing resolution can be demonstrated, obviating the need for 
separate closure models of the subgrid physics, although these may be included in a natural statistical framework.  Finally and most importantly, 
DSS leads more directly to insights, by integrating out fast modes, leaving only the slow modes that contain the 
physical information of most interest \citep{Lorenz67,Marston:2010p688}.   

In this paper we develop the techniques illustrated recently in \cite{Marston:2008p1}, where the prototypical problem of barotropic flow relaxing toward 
a point jet was considered and the statistics obtained by DNS were found to be in good qualitative agreement with 
those found from a second-order cumulant expansion. We begin by examining the general case of constructing a cumulant hierarchy for the evolution of a number of dynamic variables. We describe the derivation and solution of the cumulant equations for the general case, before focussing the discussion on the case of spherical symmetry, where computational efficiencies are available.

Having described the method in general, we illustrate the advantages of the method for
 a simple model of the 
interaction of turbulence and mean flows that may be
relevant to the generation of zonal flows in stable layers in planets
and stars. The model describes the two-dimensional evolution of flows
and magnetic fields on a spherical surface. Such an evolution is
non-trivial as it is known that for the hydrodynamic problem the
Reynolds stresses act to drive inhomogeneous zonal flows; this type of
behavior is difficult to parameterize in sub-grid scale
closures. These models and their generalizations have been used to
describe the dynamics of the outer layers of giant planets such as
Jupiter (see e.g. \cite{scottpolvani:2008} and the references therein) though competing theories for the
generation of zonal flows via deep-seated convection (see
e.g. \cite{joneskuz:2009} and the references therein) are also available. Furthermore there has been much interest in
the MHD version of this problem owing to its importance in the
dynamics of the solar tachocline (see e.g. \cite{tdh2007,hrw:2007}) 
and potentially in the outer layers of extra-solar planets
(\cite{stcho:2006}). Both of these environments are believed to be
turbulent, stably stratified and magnetized.

The tachocline is believed to  play a crucial role in the generation
of the eleven year solar cycle (see e.g.\cite{tw:2007}) and
angular momentum transport through the tachocline may be responsible
for spinning down the solar interior. One crucial issue to be resolved
is therefore the role of turbulence in transporting angular
momentum in the tachocline, and this has been addressed both in the
hydrodynamic and magnetohydrodynamic settings (see e.g. \cite{spza:1992,gm:1998,mci:2003,Tobias:2007p116}).
What has been shown is that whereas anisotropic hydrodynamic two-dimensional
turbulence leads to the efficient formation of zonal flows via 
Reynolds stresses; the addition of a magnetic field leads to Maxwell
stresses that can oppose the formation of jets. The suppression of
jets is a function of the strength of the large-scale magnetic field and the
local magnetic Reynolds number $R_m$.

The paper is organised in the following manner: in the next section we introduce the general method and the computational savings that can be achieved for the case of spherical symmetry in two dimensions. In section 3 the particular model of MHD turbulence on a rotating spherical surface is introduced and a comparison of the large-scale dynamics of DNS and DSS is made\footnote{A more in-depth discussion of the dynamics, including the calculation of turbulent transport coefficients is included in a forthcoming paper.}. We conclude by discussing extensions to the method and speculating on the range of problems where such a technique may be of use.

\section{Formulation of the Model}
\label{formulation}

In this section we describe the derivation of a general fully spectral algorithm for the direct statistical simulation of astrophysical flows \footnote{In a future paper we shall describe the adaptation of pseudospectral methods for DSS}. We develop the method for the typical case of equations with quadratic nonlinearities, before specialising to systems with spherical symmetry in two dimensions.

Consider a system that is represented by partial differential evolution equations (PDEs) for a number $r$ of scalar fields. Typically such a system may be solved directly by  discretising the PDEs using a finite-difference, finite volume or finite element method or by deriving equations for the amplitude of modes in a spectral expansion. Formally, this transforms the PDEs into a finite set of ordinary differential equations (ODEs) that may be integrated forward in time. If the discretisation is performed at $s$ discrete points (or for $s$ spectral modes) then the evolution equations can take the form
\begin{eqnarray}
\dot{q}_i &=& A_i + B_{ij}~ q_j + C_{ijk}~ q_j q_k + f_i(t)
\nonumber \\ 
&&\langle f_i(t) \rangle = 0
\nonumber \\
&&\langle f_i(t)~ f_j(t^\prime) \rangle = \Gamma_{ij}~ \delta(t - t^\prime)\ .
\label{algebraicEOMs}
\end{eqnarray}
where $1 \le i,~j,~k \le r s$  and $A_i$,  $B_{ij}$ and $C_{ijk}$ are the coefficients. Here the  $q_i$ are the discretised values of the dependent variables (or the amplitudes of the relevant spectral modes); typically these represent a vector of the values of the fluid properties. We also note here that there is an implicit sum over repeated indices. 

Hereinafter, to fix ideas, we shall think of the $q_i$ as representing the amplitudes of the spectral modes of a vector of dependent variables --- and shall give a concrete example in the next section. The forcing $f_i(t)$ can then be interpreted as the statistical forcing of the relevant spectral mode.

\subsection{The general case}\label{ce}

\subsubsection{Reynolds decompositions, cumulants and moments}

One way to formulate the cumulant expansion is by carrying out a Reynolds decomposition of the dynamical variable $q_i$ into the sum of a mean value and a fluctuation (or eddy):
\begin{eqnarray}
q_i = \langle q_i \rangle + q_i^\prime\  \ {\rm with}~ \langle q_i^\prime \rangle = 0 
\label{ReynoldsDecomposition}
\end{eqnarray}
where we defer, for now, choosing the type of averaging denoted by the angular brackets $\langle \rangle$.  Typical choices are  temporal or zonal averages,\footnote{Although the $q_i$ are functions of time only, a zonal average may be taken by  keeping only those modes that correspond to axisymmetric modes (for an expansion in spherical
basis functions this is equivalent to keeping the $m=0$ modes).} or averages over an ensemble of initial conditions or an ensemble of realisations. 

Once the Reynolds decomposition has been implemented, progress is made by defining the first three equal-time cumulants $c_i$, $c_{ij}$, and $c_{ijk}$ of the combined scalar fields ($q_i$) as \footnote{These definitions are sufficient for cumulant hierarchies truncated at either second or third order; for higher order hierarchies corresponding definitions of the higher order cumulants are required.}: 
\begin{eqnarray}
c_i &\equiv& \langle q_i \rangle = m_i,
\nonumber \\ 
c_{ij} &\equiv& \langle q_i^\prime q_j^\prime \rangle ,
\nonumber \\ 
&=& \langle q_i q_j \rangle - \langle q_i \rangle \langle q_j \rangle = m_{ij} - m_i~ m_j,
\nonumber \\
c_{ijk} &\equiv& \langle q_i^\prime q_j^\prime q_k^\prime  \rangle,
\nonumber \\
&=& \langle q_i q_j q_k \rangle - (\langle q_i \rangle \langle q_j q_k \rangle + perms) 
+ 2 \langle q_i \rangle \langle q_j \rangle \langle q_k \rangle,
\nonumber \\
&=& m_{ijk} - m_i m_{jk} - m_j m_{ik} - m_k m_{ij} + 2 m_i m_j m_k,
\label{cumulants}
\end{eqnarray}
where $m_i$, $m_{ij}$, and $m_{ijk}$ are respectively the traditional definitions of the first, second and third moments.  
We stress here that the second and higher cumulants  contain information about correlations that are non-local in space and therefore include interactions that are not included in the simple local moment hierarchies discussed in the introduction. For this reason this approach is more tailored to inhomogeneous problems.  

\subsubsection{Derivation of the cumulant hierarchy: the Hopf functional approach}

The hierarchy of equations of motions for the evolution of the cumulants can 
 be obtained directly be differentiating Eqs. \ref{cumulants} with respect to time and using Eqs. \ref{algebraicEOMs}, together with repeated back substitution. A more elegant method is to introduce variables $p_i$ that are, in analogy to quantum mechanics, conjugate to the $q_i$ in the sense that $q_i = -i \partial / \partial p_i$ as in Eq. \ref{Hopf-moments} below \citep{ma:2005p108}.   Then one may define the Hopf generating functional \citep{frisch95}:
\begin{eqnarray}
\Psi[q(t),~ p] &\equiv& e^{i p_i q_i(t)},
\label{Psi}
\end{eqnarray}
recalling the summation over repeated indices.
The Hopf functional obeys a Schr\"odinger-like equation:
\begin{eqnarray}
i \frac{\partial}{\partial t} \Psi &=& \hat{H} \Psi,
\label{HopfEquation}
\end{eqnarray}
with linear operator $\hat{H}$ given by:
\begin{eqnarray}
\hat{H} &\equiv& p_i \left(-A_i + i B_{ij}~ \frac{\partial}{\partial p_j} + C_{ijk} \frac{\partial^2}{\partial p_j \partial p_k} \right),
\label{linearOp1}
\end{eqnarray}
as can be verified by combining Eqs. \ref{Psi}, \ref{HopfEquation}, and \ref{linearOp1} to reproduce Eq. \ref{algebraicEOMs} in the absence of any stochastic forcing. 

As Eq. \ref{HopfEquation} is linear in $\Psi$, the average $\overline{\Psi} \equiv \langle \Psi[q(t),~ p]  \rangle$ obeys the same equation; however $\overline{\Psi}$ encapsulates information about the equal-time moments, as can be seen by repeated differentiation of Eq. \ref{Psi} with respect to $p_i$, followed by averaging:
\begin{eqnarray}
\langle q_{i_1} q_{i_2} \cdots q_{i_n} \rangle = (-i)^n~ \frac{\partial^n \overline{\Psi}}{\partial p_{i_1} \partial p_{i_2} \cdots 
\partial p_{i_n}} \bigg{|}_{p_i = 0}.
\label{Hopf-moments}
\end{eqnarray}
(For the case of time-averaging, the statistics do not vary in time, $\frac{\partial}{\partial t} \overline{\Psi} = 0$, and the statistics are obtained from the solution of the time-independent equation $\hat{H} \overline{\Psi} = 0$.)
The Hopf functional $\overline{\Psi}$ may also expressed as the exponential of a power series in $p_i$, the coefficients being the cumulants:
\begin{eqnarray}
\overline{\Psi} = \exp \left\{ i c_i(t)~ p_i - \frac{1}{2!} c_{ij}(t)~ p_i p_j - \frac{i}{3!} c_{ijk}(t)~ p_i p_j p_k + \ldots \right\}
\label{Hopf-cumulants}
\end{eqnarray}
as can be checked by use of Eq. \ref{Hopf-moments} to reproduce the moments in terms of the cumulants, Eqs. \ref{cumulants}.  
Stochastic forcing can now be included with the addition of the $\Gamma_{ij}$ term:
\begin{eqnarray}
\hat{H} &\equiv& p_i \left(-A_i + i B_{ij}~ \frac{\partial}{\partial p_j} + C_{ijk} \frac{\partial^2}{\partial p_j \partial p_k} + i \Gamma_{ij} p_j \right)\ .
\label{linearOp2}
\end{eqnarray}
Upon substituting Eq. \ref{Hopf-cumulants} into Eq. \ref{HopfEquation} and collecting powers of $p_i$ one obtains the equations of motion (EOM) for the cumulants that truncated at third order read 
\begin{eqnarray}
\dot{c}_i &=& A_i + B_{ij}~ c_j + C_{ijk}~ (c_j~ c_k + c_{jk})
\nonumber \\
\dot{c}_{ij} &=& \{2 B_{ik}~ c_{kj} + C_{ik\ell}~ (4 c_\ell~ c_{jk} + 2 c_{jk\ell})\} + \Gamma_{ij}
\nonumber \\
\dot{c}_{ijk} &=& \{3 B_{i \ell}~ c_{\ell j k} + 6 C_{k \ell m}~ (c_{ijm}~ c_\ell + c_{im}~ c_{j \ell})\} - \mu~  c_{ijk} + {\cal O}(c_{ijk\ell})\ 
\label{ce-eom}
\end{eqnarray}
where we defer discussion of the parameter $\mu$ until later.
Here for compactness we have introduced the short-hand notation $\{ \}$ to denote symmetrization over indices
\begin{eqnarray}
\{2 B_{ik}~ c_{kj}\} \equiv  B_{ik}~ c_{kj} + B_{jk}~ c_{ki}
\end{eqnarray}
that maintains symmetries $\dot{c}_{ij} = \dot{c}_{ji}$ and similarly for the third cumulant.  

Truncated at second-order (CE2) the cumulant expansion is realizable \citep{salmon98}\footnote{CE2 can be viewed as the exact solution of a stochastically-driven linear model} and well-behaved in the sense that the energy density is positive and the second cumulant obeys positivity constraints.  Going to third-order (CE3) and beyond introduces difficulties.  A phenomenological eddy-damping parameter \citep{Orszag:1977,Andre:1974p398} $\mu$ that models the neglect of the fourth and higher cumulants from the hierarchy is included in the last of Eq. \ref{ce-eom} and is required to prevent blow-up.  This ad-hoc procedure is somewhat unsatisfactory and more robust methods may be necessary. Indeed determining reliable methods of truncating the hierarchy is a matter of current research.

\subsection{Symmetry and the derivation of reduced cumulant equations.}
\label{symmetries}

In principle, the general set of cumulant equations in Eq.~\ref{ce-eom} can be solved with enough computational effort. However, efficient algorithms can be developed if the underlying system exhibits further symmetries. This is typically the case for astrophysical systems, which usually exhibit spherical or cylindrical symmetry or a corresponding translational symmetry in a local Cartesian domain. We discuss in detail here the case of cumulants in a sphere.

\subsubsection{Equations for fully spectral DNS on a sphere}

For systems with an underlying spherical symmetry, the spectral expansion of the dependent variables discussed in section~\ref{formulation} often takes the form
\begin{eqnarray}
q &=& \sum_{\ell; m}^{L; M} q_{\ell m}(r) Y_\ell^m(\theta,\phi) , 
\nonumber \\
&=& \sum_{\ell; m}^{L; M} q_{\ell m}(r)~ (-1)^m \sqrt{\frac{2 \ell + 1}{4 \pi} \frac{(\ell - |m|)!}{(\ell + |m|)!}}~ P_\ell^m(\cos \theta) e^{i m \phi},
\end{eqnarray}
where $r$ is spherical radius, $\theta$ is co-latitude and $\phi$ is longitude. Here the $q_{\ell m}(r)$ are complex functions and the $P_\ell^m$ are associated Legendre functions. Furthermore on a spherical surface the $r$-dependence is absent and a fully spectral representation of the equation of motion (equation~\ref{algebraicEOMs}) can be written as
\begin{eqnarray}
\dot{q}_{\ell m} &=& A_{\ell}~ \delta_{m,0} + \sum_{\ell_1} B_{\ell; \ell_1 m}~ q_{\ell_1 m} + f_{\ell m}(t)
\nonumber \\
&+& \sum_{\ell_1, \ell_2, m_1, m_2}^{m = m_1+m_2} C^{(+)}_{\ell; \ell_1 m_1; \ell_2 m_2}~  q_{\ell_1 m_1} q_{\ell_2 m_2} 
\nonumber \\
&+&  \sum_{\ell_1, \ell_2, m_1, m_2}^{m = m_1-m_2} C^{(-)}_{\ell; \ell_1 m_1; \ell_2 m_2}~  q_{\ell_1 m_1} q^*_{\ell_2 m_2}\ .
\label{spectralEOM}
\end{eqnarray}
We note here that, because the scalar fields are real-valued in coordinate space, we may focus on the evolution of modes with $m \geq 0$ as modes with $m < 0$ may be obtained by complex conjugation. Moreover, for simplicity in the above and in subsequent equations the index that encodes which state variable is being solved for has been subsumed into the $\ell$ label.  

The quadratic nonlinearities have their origin in the Jacobians of Eqs. \ref{EOM} with coefficients $C^{(+)}$ representing amplitudes for the scattering of two waves with $m \geq 0$; $C^{(-)}$ are for waves with $m > 0$ and $m < 0$ to scatter.   The amplitudes of these coefficients are constructed from the matrix elements of the Jacobian:
\begin{eqnarray}
I^{(\pm)}_{\ell; \ell_1 m_1; \ell_2 m_2} \equiv i m_1 \int d\theta~ P_{\ell}^{m}(\cos \theta)~ P_{\ell_1}^{m_1}(\cos \theta)~ \frac{\partial}{\partial \theta} P_{\ell_2}^{m_2}(\cos \theta)
\end{eqnarray}
where $m = m_1 \pm m_2$.  Integrals $I^{(\pm)}$ are evaluated in a numerically exact manner by Gaussian quadrature.

\subsubsection{Equations for fully spectral DSS on a sphere}

Similar considerations lead to an efficient representation of the cumulant hierarchy for a spherical shell. These considerations can then be combined with the knowledge of the underlying symmetries of the statistics themselves to derive reduced hierarchies of cumulant equations. These symmetries are preserved whether zonal, temporal or ensemble averages are used.
Statistics on the rotating sphere exhibit azimuthal symmetry.   The simplest conceptual choice for the averaging operation $\langle \rangle$ is therefore the zonal average and we choose that here, and then follow that with a running time average.  On symmetry grounds, the first cumulant must be independent of longitude $\phi$ and therefore in the spherical harmonic basis only the $m = 0$ mode $c_\ell = \langle q_{\ell, m=0} \rangle$ is non-zero.  Similar symmetry arguments yield the result that the second cumulant depends on the latitudes of the two field points, but only on the {\em difference between} their longitudes.  It can therefore be written as $c_{\ell_1 \ell_2 m} = \langle q_{\ell_1 m} q_{\ell_2 -m} \rangle - c_{\ell_1} c_{\ell_2} \delta_{m 0}$.  Furthermore, zonal averaging then requires that $c_{\ell_1 \ell_2 m=0} = 0$.  Similarly the third cumulant is a function of only 5, not 6, wavenumbers, i.e.\ it can be written as  $c_{\ell; \ell_1 m_1; \ell_2 m_2}$.  Moreover,
because the scalar fields are real-valued in coordinate space, we have $c_{\ell_1 \ell_2 m} = c_{\ell_2 \ell_1 m}^*$.  For models with an imposed north-south reflection symmetry about the equator\footnote{this is not necessary, but is computationally expedient}, the cumulants respect further constraints:  $c_\ell$ vanishes for all even $\ell$ and $c_{\ell_1 \ell_2 m} = 0$ if $\ell_1$ is odd and $\ell_2$ is even, and vice-versa.  All of these symmetries therefore lead to a computational saving.

We consider here the simplest non-trivial case where the hierarchy is truncated at second order (CE2), i.e.\ all higher cumulants are set to zero.  The EOM for the cumulants in the basis of spherical harmonics are then
\begin{eqnarray}
\dot{c}_\ell &=& A_\ell + B_{\ell; \ell_1 0}~ c_{\ell_1} + C^{(-)}_{\ell; \ell_1 m; \ell_2 m}~  c_{\ell_1 \ell_2 m},
\nonumber \\
\dot{c}_{\ell_1 \ell_2 m} &=& 2 \Gamma_{\ell_1 m} \delta_{\ell_1 \ell_2} + B_{\ell_1; \ell m}~ c_{\ell \ell_2 m} + B_{\ell_2; \ell m}~ c_{\ell_1 \ell m}
\nonumber \\
&+& C^{(+)}_{\ell_1; \ell 0; \ell^\prime m}~ c_{\ell}~ c_{\ell^\prime \ell_2 m} 
+ C^{(+)}_{\ell_2; \ell 0; \ell^\prime m}~ c_{\ell}~ c_{\ell_1 \ell^\prime m},
\label{ceom2}
\end{eqnarray}
where again the convention of summation over repeated indices has been adopted.  (There would also be a contribution to the first cumulant from $C^{(+)}_{\ell_1; \ell 0; \ell^\prime 0}~ c_{\ell}~ c_{\ell^\prime}$ but it vanishes for the problems considered here as the Jacobian of two fields with no longitudinal dependence is zero.)  These equations may be compared to a coordinate-independent version given by Equations~(21) and (22) in \cite{Marston:2008p1}.  That only the eddy-mean flow interaction is retained in CE2 may be seen by noting that the coupling of the first cumulant with the second involves no mixing of the azimuthal wavenumber $m$ (only a single $m$ appears in Eqs. \ref{ceom2}).   Eddy-eddy scattering occurs only at third and higher orders\footnote{We shall investigate including higher orders in the hierarchy for the cumulant expansion in subsequent papers.}.

\section{Turbulent driven MHD on a Spherical Surface: The Model}
\label{model}

We consider a simple two-dimensional model of a stably stratified
region of hydrodynamic or MHD turbulence. This is the simplest extension of the local $\beta$-plane model considered by \cite{Tobias:2007p116}. We stress again that, although this system is of interest in its own right and the interaction of Reynolds and Maxwell stresses play an important role in the dynamics of the tachocline and other regimes of stably stratified MHD turbulence, in this paper we are utilising this model as a non-trivial example of the utility of direct statistical simulation. We therefore defer discussion of the interaction of the stresses for a subsequent paper.

The behaviour of such a
system in two dimensions can be described by the evolution of two
scalar fields, namely the relative vorticity $\zeta(\theta, \phi, t)$ (with
$\theta$ being co-latitude and $\phi$ longitude, as before) and the scalar potential
for the magnetic field $A(\theta, \phi, t)$ \citep[cf][]{Tobias:2007p116}. When
extended to the sphere rotating at angular rate $\Omega$ these may be written:
\begin{eqnarray}
\dot{q}  &=& J[q,~ \psi] + J[A, \nabla^2 A^\prime]  -  \kappa~ \zeta - \nu_2 \nabla^4 \zeta + f(t),
\nonumber \\ 
\dot{A}  &=& J[A,~ \psi]  + \eta \nabla^2 A^\prime,
\label{EOM}
\end{eqnarray}
where on the unit sphere the Jacobian is given by
\begin{eqnarray}
J[q,~ \psi] \equiv \frac{1}{\sin \theta} 
  \left(\frac{\partial q}{\partial \phi} 
    \frac{\partial \psi}{\partial \theta} -
    \frac{\partial q}{\partial \theta} 
    \frac{\partial \psi}{\partial \phi} \right). 
  \label{Jacobian}
\end{eqnarray}
Here
\begin{eqnarray}
\zeta &=& \nabla^2 \psi,
\nonumber \\
q &=& \zeta + 2 \Omega \cos \theta, 
\nonumber \\
A &=& A^\prime +  B_0 \cos \theta.
\end{eqnarray}
Hence $q$ is the absolute vorticity. Here $\kappa$ is a frictional term,
$\nu_2$ is a hyperviscosity whilst  $f(t)$ is the stochastic
forcing. Magnetic diffusion is explicitly included through a magnetic
diffusivity  $\eta$, since we believe it is important to capture this
process correctly.
We note here that the parameter $B_0$ measures the strength of a 
toroidal imposed magnetic field, which is held fixed in time and note that such a
field can not be self-consistently maintained in a strictly
two-dimensional calculation. The equations have been scaled so that
the magnetic field is measured in units of the Alfv\'en velocity. For
purely hydrodynamic simulations we simply set $B_0 = A = 0$.

These equations may be then be written in the form of equation~(\ref{algebraicEOMs}) (with $r=2$)
by setting  the absolute
vorticity and magnetic potential scalar fields into two layers
labelled by $q_\alpha$ with $q_1 = q$ and $q_2 = A$ and discretising the system either to obtain  equations for the spectral amplitudes of the form equation~(\ref{spectralEOM}) or more conveniently for computation a finite difference representation on a spherical geodesic grid. In a similar manner the spectral representation of the cumulant equations (i.e.\ equations~(\ref{ceom2})) can simply be derived once the coefficients in equation~(\ref{spectralEOM}) have been calculated for this model.

\section{Comparison of DNS and DSS}
\label{comparison}

\subsection{Numerical implementation of DNS and DSS}

\subsubsection{DNS}

Direct numerical simulation of the two-dimensional system has been implemented using two different techniques. The EOM given by equation~(\ref{spectralEOM}) may be integrated forward in time in their pure spectral form using a standard fourth-order accurate Runge-Kutta algorithm with an adaptive time step, though in practice it is much faster  work directly in real space on a spherical geodesic grid as we do here. The fully spectral code is therefore only used as a validation of the geodesic code below and a useful comparison with the fully spectral direct statistical simulation.

The most efficient numerical
integration of the DNS EOM is carried out in real space on a spherical
geodesic grid \citep{Heikes:1995p113} of $D$ cells with the use of the
second-order accurate leapfrog algorithm and a Robert filter.  A
multigrid algorithm solves Poisson's equation at each time step.

\subsubsection{DSS}

We take advantage of the stiff nature of the spectral EOM for the cumulants (equations~\ref{ceom2}).  These are integrated forward in time using a semi-implicit backward Euler Full Orthogonal Method \citep{Saad:2003} that is based upon Krylov subspaces and that permits a much longer time step than is possible for explicit integration methods. 

We note here that integration of the EOM for CE2, equations~(\ref{ceom2}), requires of order $L^3 M$ operations at each time step, where $0 \leq \ell \leq L$ and 
$0 \leq m \leq {\rm min}(\ell, M)$ define the spectral cutoffs.  A pseudo-spectral implementation of the EOM would require the same order of operations on the sphere and thus offers no advantages over the pure spectral method used here.    We find that all $c_{\ell_1 \ell_2 m}$ with $m$ greater than the maximum azimuthal wavevector of the stochastic forcing vanish, hence the spectral expansion can be severely truncated by restricting $M \ll L$ without loss of accuracy. This results in substantial speed-up and a reduction in the required memory. Moreover, only a subset of the possible coefficients of the quadratic nonlinearities, $C^{(-)}_{\ell; \ell_1 m; \ell_2 m}$ and $C^{(+)}_{\ell_1; \ell 0; \ell^\prime m}$, with 4 indices appear in Eq. \ref{ceom2} resulting in reduced memory usage.

Finally we note that the code implementing both DNS and DSS (via CE2) is written in the Objective-C++ programming language and runs on Apple computers (OS X 10.6) utilizing C-blocks and grand central dispatch (gcd) for efficient SMP parallelism. We stress that the DSS can run an order of magnitude or more faster than DNS.

\subsection{Conservation Laws, Model Parameters and Initial Conditions}
\label{parameters}

In the absence of damping and driving forces, the EOM for the cumulants, like the EOM for the vorticity and magnetic potential have a number of conservation laws. For example, in the hydrodynamic case, kinetic energy, enstrophy and angular momentum are conserved, whilst for the MHD case the conserved quantities are 
angular momentum, total energy, cross-helicity, and the mean squared potential.  Moreover, for stochastic forcing restricted to wavevectors $| \ell | > 0$, the case considered here, the angular momentum in the CE2 remains exactly zero, in contrast to DNS.

Just as for direct numerical simulations utilising spherical harmonics there are convenient expressions of the average values of various quantities
in terms of the low-order cumulants.  For example the mean cross-helicity is given by:
\begin{eqnarray}
\frac{1}{4 \pi} \int d^2\Omega~ \langle \vec{v} \cdot \vec{B} \rangle 
&=& - \frac{1}{4 \pi} \int d^2\Omega~ \langle q A \rangle
\nonumber \\
&=& - \frac{1}{2 \pi} \sum_{\ell m} (c_{1\ell 2\ell m} + c_{1\ell}~ c_{2\ell}~ \delta_{m0} ) 
\end{eqnarray}
where the two layers are labelled explicitly in the final line.  Similar expressions are available for the averages of other quadratic quantities.

The models are formulated on the unit sphere with a timescale such that the sphere complete a full rotation in one day of model time.   All model parameters may be defined in terms of these length and time scales; for instance $\Omega = 2 \pi$.  Friction removes energy at long length scales and is parameterized by rate $\kappa$.  
The hyperviscosity $\nu_2$ that appears in Eq.~(\ref{EOM}) is included solely to absorb enstrophy at the smallest resolved scales.  Consequently it is rescaled with the grid size or spectral cutoff so that
\begin{eqnarray}
\nu_2 &\rightarrow& \nu_2 * (D/4)^{-2} \ \ {\rm geodesic},
\nonumber \\
\nu_2 &\rightarrow& \nu_2 * (L (L+1))^{-2} \ \ {\rm spectral},
\end{eqnarray}
where the maximum eigenvalue of $-\nabla^2$ on a geodesic grid with $D$ cells is approximately $D/2$.     Thus for $\nu_2 = 1$ (the case we consider here) features on the smallest length scales are dissipated on a time scale of order $1$ day. 

Stochastic forcing is confined to wavevectors $L_{min} \leq \ell \leq L_{max}$ and $M_{min} < | m | \leq L_{max}$.  Within this range of wavevectors the forcing $f_{\ell m}$ that appears in 
Eq.~(\ref{spectralEOM}) is given by
\begin{eqnarray}
f_{\ell m}(t) = \sqrt{F / \Delta} * {\rm gaussian}(t / \Delta)
\end{eqnarray}
where ${\rm gaussian}(t)$ is a complex number randomly drawn, for each value of $\ell$ and $m$, from a normal distribution of zero mean and unit variance that smoothly transitions from one random number to the next over a time period of $\Delta$.  We set $\Delta = 0.1$ which is large compared with the time step, but small compared with advective time scales.  Consequently in Eq.~(\ref{ceom2}) we have
\begin{eqnarray}
\Gamma_{\ell m} =  \begin{cases}
2 F &\,\,for \,\, L_{min} \leq \ell \leq L_{max}\,\, and \,\, M_{min} \leq | m | \leq \ell \\
0 & \,\,else.
\end{cases}
\end{eqnarray}
In the following we hold fixed $\kappa = 0.02$, $\nu_2 = 1$,  $F = 0.2$, and (for the magnetic cases) $\eta = 10^{-4}$. We study the evolution of the systems (DNS and DSS)  for two different choices for the range of the forcing wavevectors $\{L_{min},M_{min},L_{max}\}$.

We close our description of the set-up of the models by commenting on the choice of initial condition. The DNS integrations are started from rest with zero perturbation to the imposed field. 
 For the DSS, at the start of the CE2 integration we set the first
 cumulant $c_\ell = 0$ and the second cumulant $c_{\ell_1 \ell_2 m} =
 c_2~ \delta_{\ell_1 \ell_2}$ which corresponds to initial
 short-ranged correlations in the vorticity.  At low resolutions, the
 fixed point sometimes has jets that move in directions opposite to
 those found in DNS; this fixed point, which is an artifact of the spectral truncation,
 can be avoided by initializing $c_\ell$ with small values.

\section{Results}

\begin{table}%[H] add [H] placement to break table across pages
 \begin{center}
 %\begin{table}
 \begin{tabular}{|   c |   c |  c |  c |  c |  c |  }
 \hline
 Case & Method & $B_0$ & $Ro$ &  $Rm$ &  $\langle B^2 \rangle / \langle U^2 \rangle$ \\ \hline
 1 & DNS & 0 &  0.0351&  4415  &   0.0\\ \hline
 1 & DSS(CE2) & 0 &  0.0352&  4431  &  0.0\\ \hline
 1 & DNS & 0.1 &  0.0331&  4195  &   0.04\\ \hline
 1 & DSS(CE2) & 0.1 &  0.0323&  4064 &    0.04\\ \hline
 1 & DNS & 0.5 &  0.015&   1865 & 1.29 \\ \hline
 1 & DSS(CE2) & 0.5 & 0.017 & 2144 &   1.02\\ \hline \hline
 2 & DNS & 0 &  0.0542 &  6812 &    0.0\\ \hline
 2 & DSS(CE2) &  0 & 0.0548 & 6885   &   0.0 \\ \hline
 2 & DNS & 1.0 & 0.0237 & 2980  & 1.07 \\ \hline
 2 & DSS(CE2) & 1.0 & 0.0396 & 4980  & 1.25  \\ \hline
 \end{tabular}
\end{center}
\caption{Non-dimensional numbers as calculated a posteriori. Case 1 refers to stochastic forcing with $M_{min}=8$ whilst Case 2 refers to $M_{min}=1$.  
(For the case of zero magnetic field, $Rm$ reduces to a non-dimensional measure of the kinetic energy of the flow.)
\label{nondim-numbers}}
 %\end{table}
 \end{table}

\subsection{Small-scale forcing:  $L_{min} = 8$, $M_{min} = 8$, and $L_{max} = 12$}

\begin{figure}
\includegraphics[width=3in,height=1.8in]{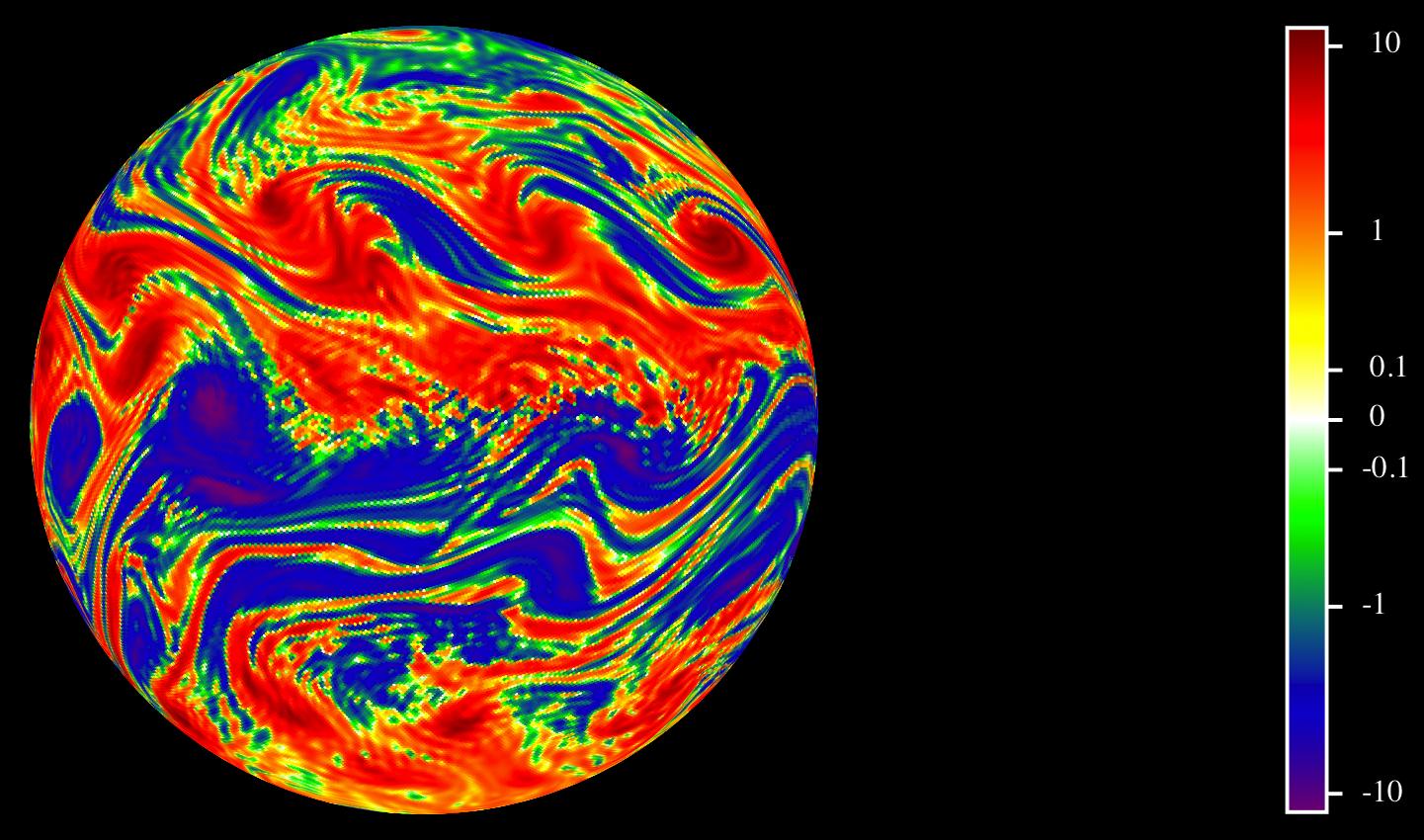}
\includegraphics[width=3in,height=1.8in]{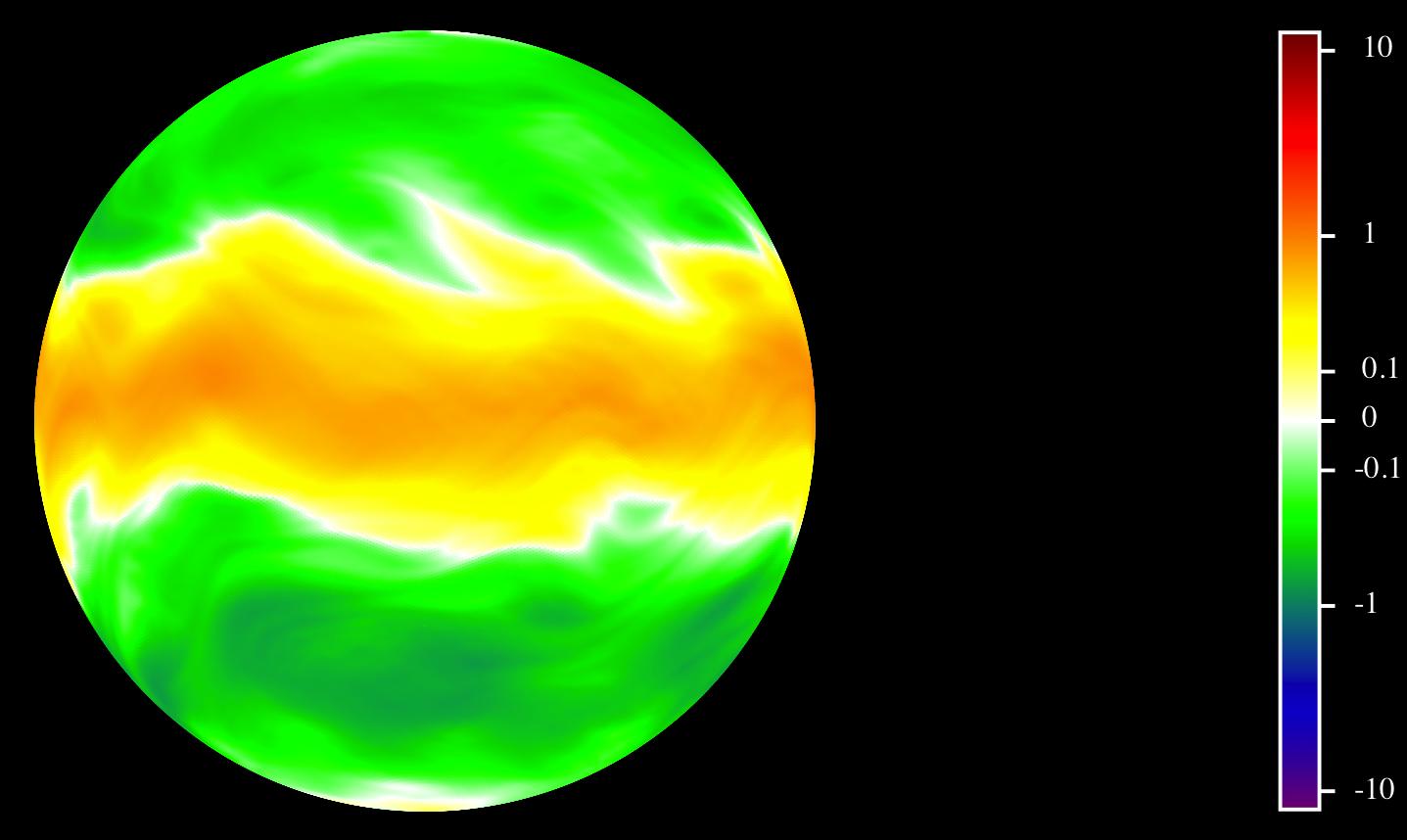}
\vskip 0.05in
\includegraphics[width=3in,height=1.8in]{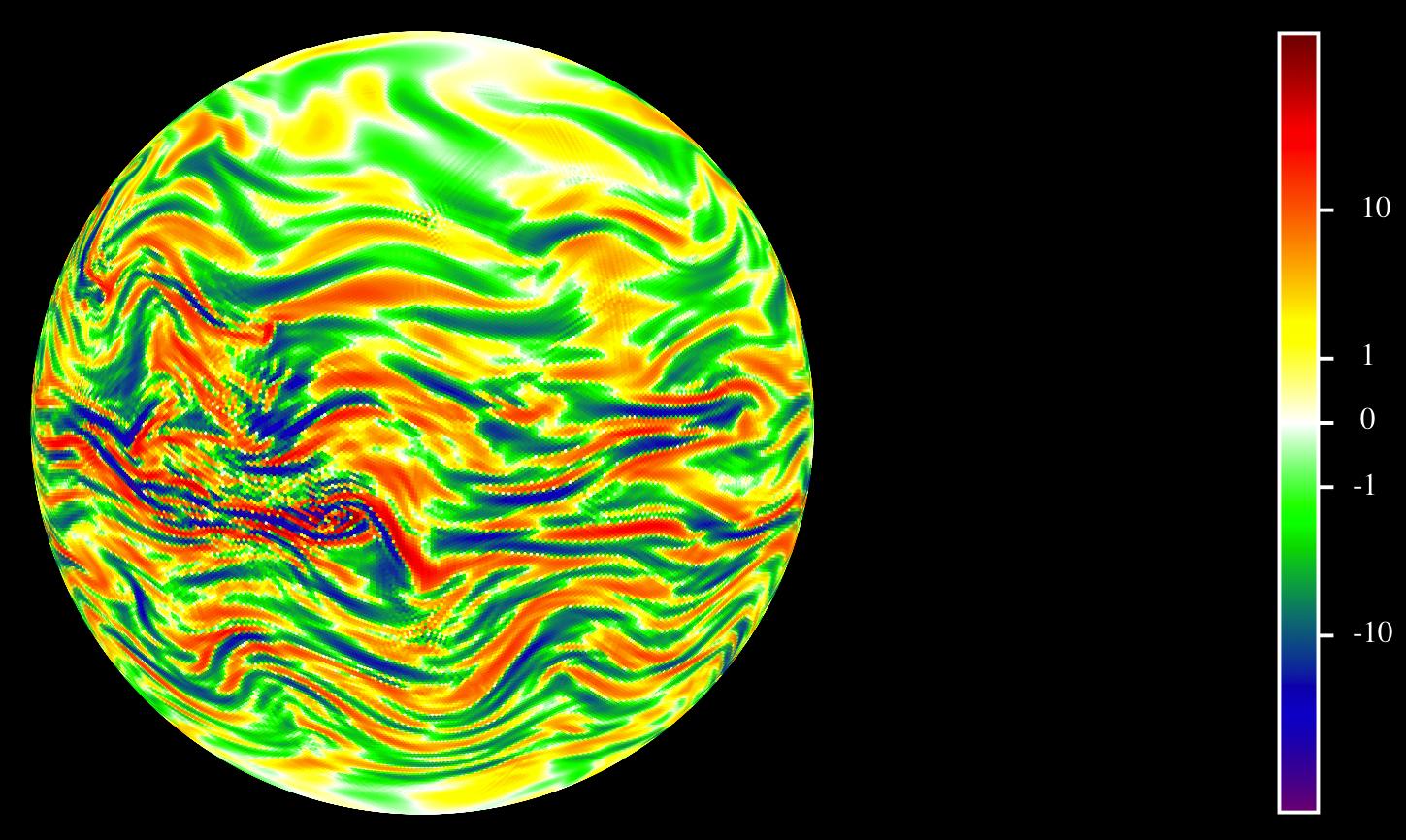}
\includegraphics[width=3in,height=1.8in]{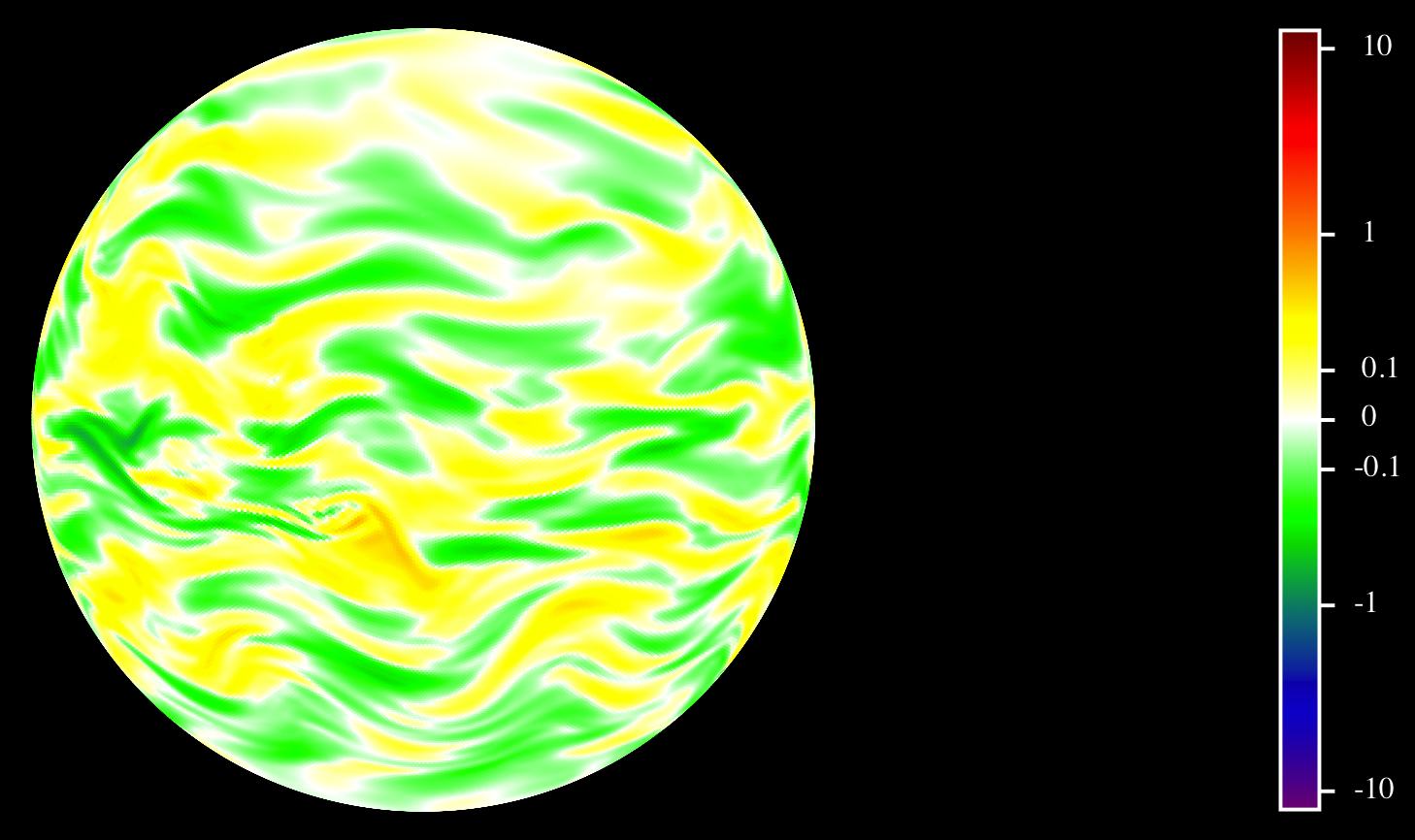}
\caption{DNS calculation of the instantaneous relative vorticity (left panels) and zonal velocity fields (right).  Top: Pure hydrodynamic model ($B_0 = 0$) that develops a westerly (prograde) jet along the equator.  Bottom: Imposed toroidal magnetic field parameter $B_0 = 0.5$.  The calculation is done on a spherical geodesic grid with $D = 163,842$ cells with stochastic forcing over spherical wavevectors $8 \leq \ell,~ m \leq 12$.  See Sec. \ref{parameters} for the values of the other parameters.}
\label{figure1}
\end{figure}

We begin by considering the hydrodynamic and magnetohydrodynamic
evolutions for the case where the system is forced solely at small
scales in the vorticity equation. The  DNS of the hydrodynamic case is
performed until a statistically steady state is reached and meaningful
statistics can be calculated. For this case, this has occurred by $t
\sim 1000$; after this time a running time-average is performed for
another $1000$ days. Here the small-scale driving leads to the
formation of flows on a range of scales including large-scale jets as
shown in Figure~\ref{figure1} (top panels) which show the
instantaneous relative vorticity (left) and relative zonal velocity
(right). These clearly show the formation of  a prograde (westerly)
jet at the equator with two retrograde (easterly) jets at high
latitudes, with the total angular momentum of the fluid remaining
close to zero. As we shall see, these jets are driven by the flows on
smaller scales. 

\begin{figure}
\includegraphics[width=3in]{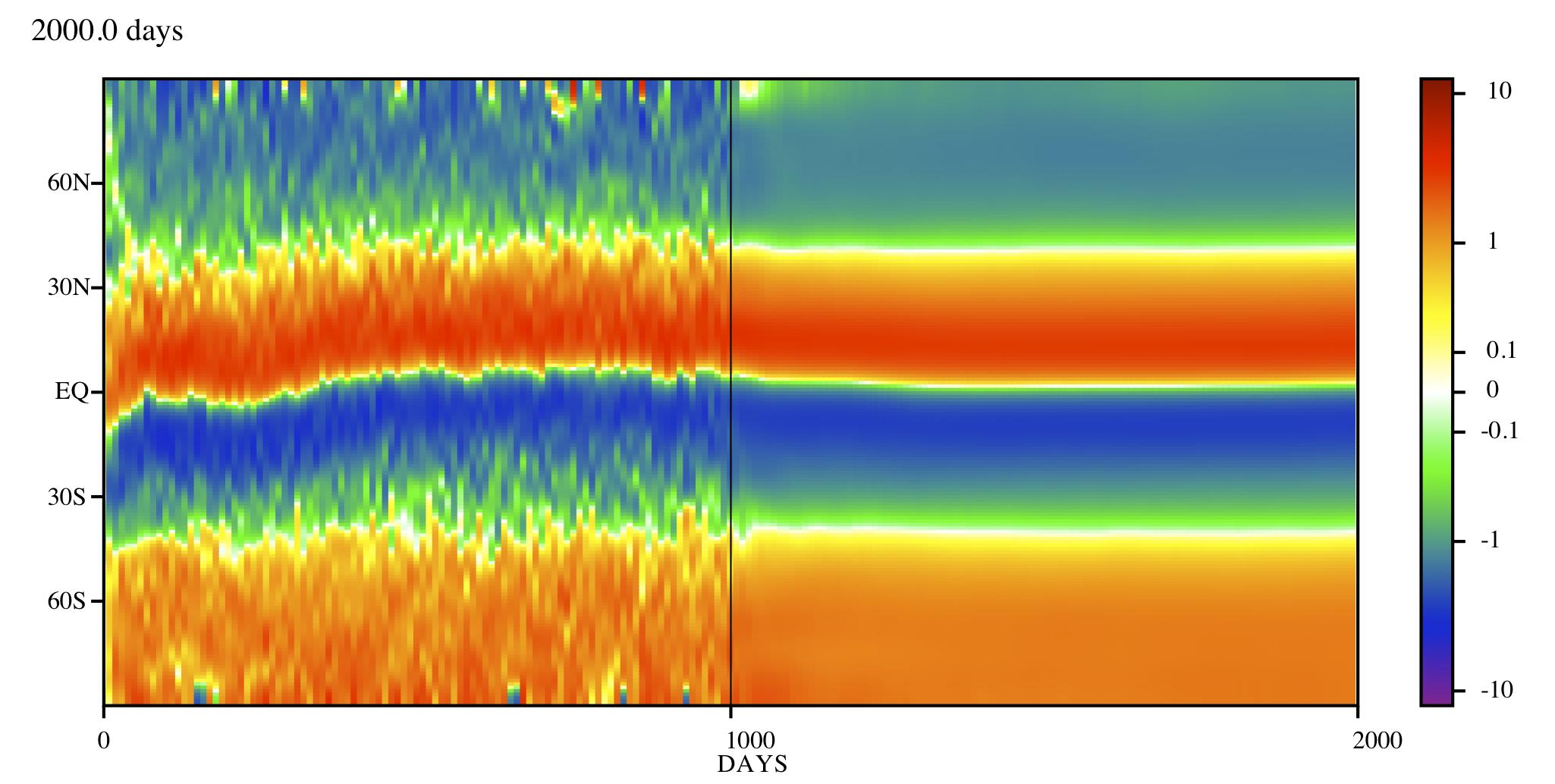}
\includegraphics[width=3in]{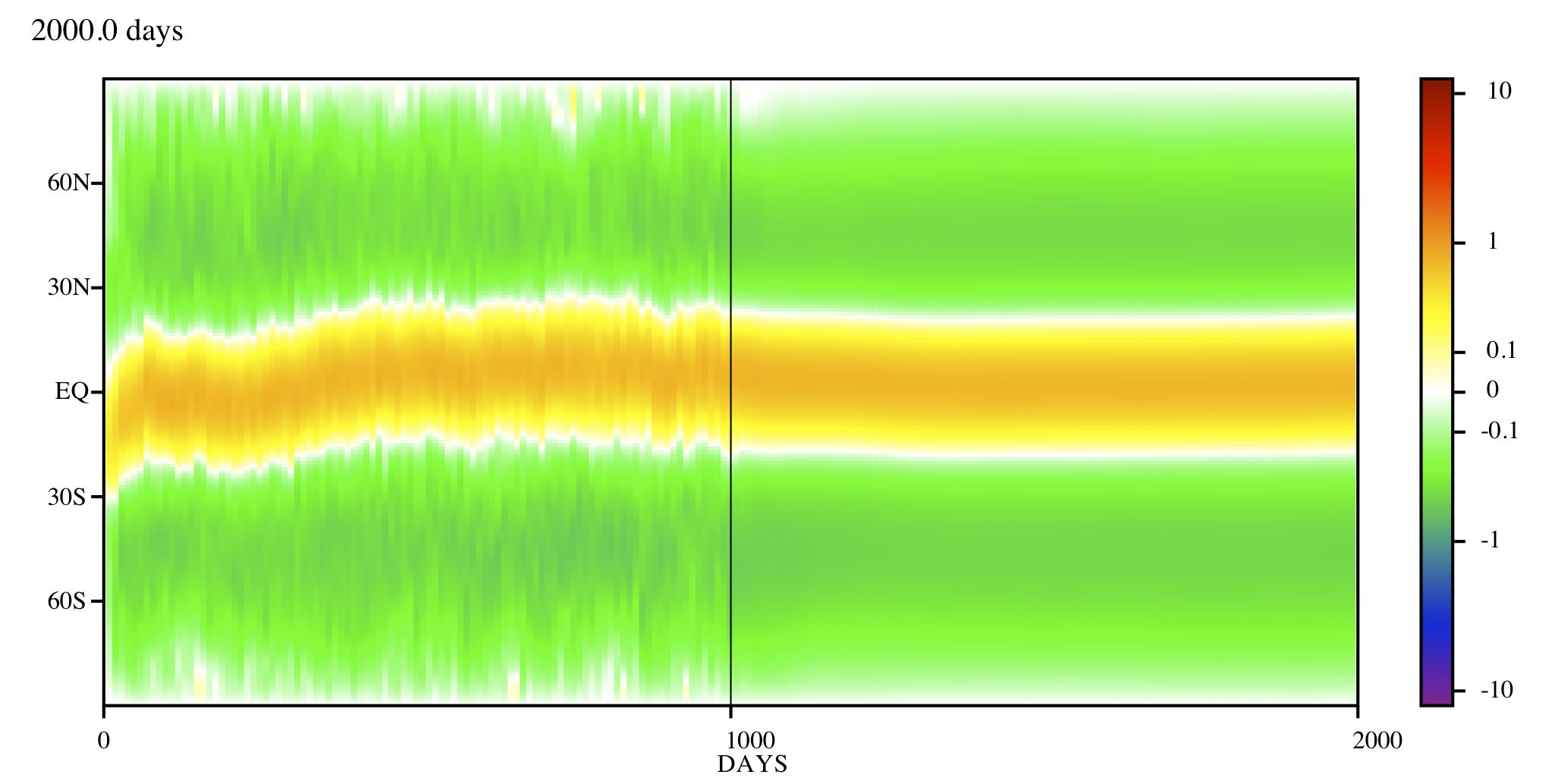}

\includegraphics[width=3in]{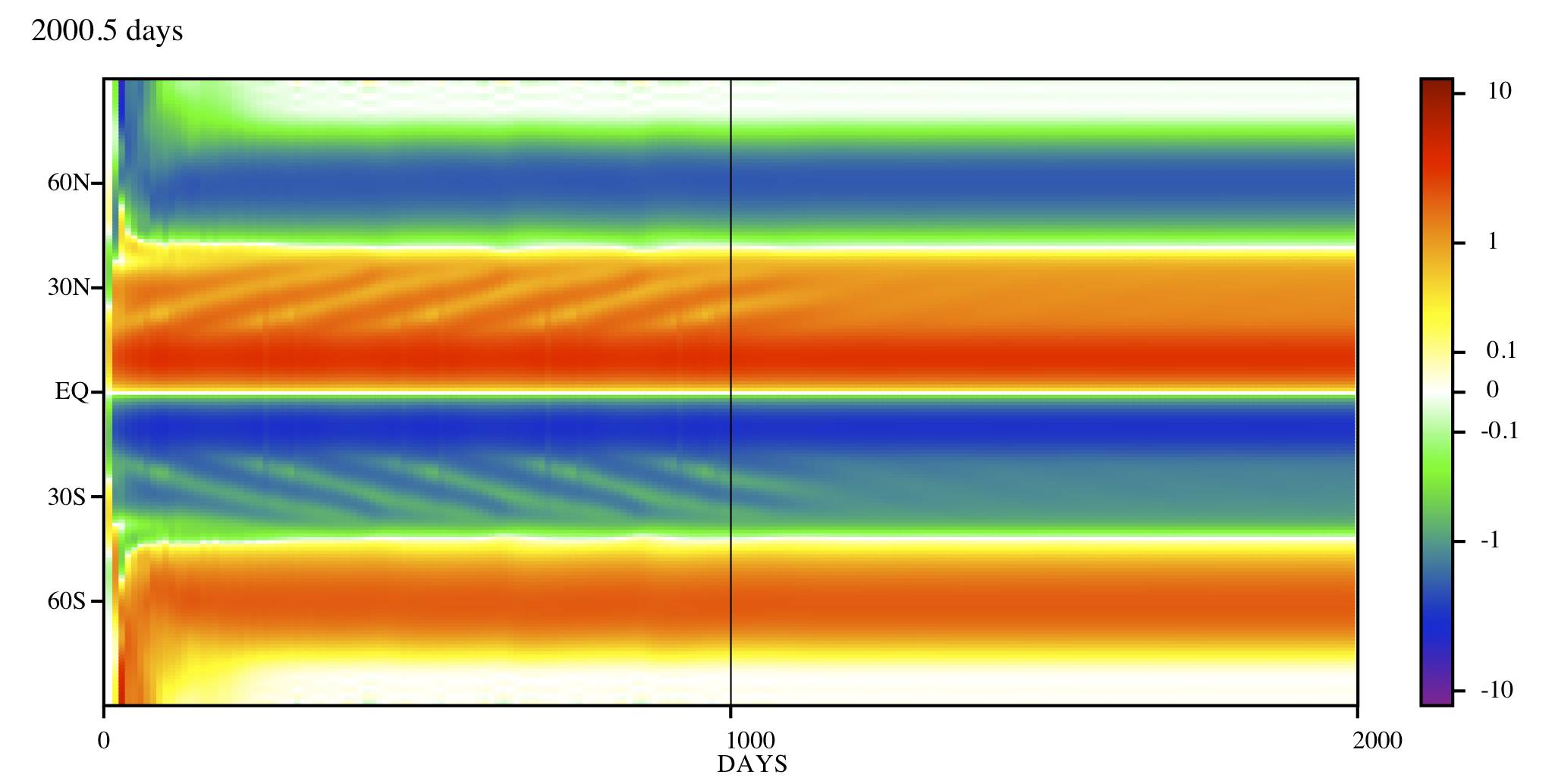}
\includegraphics[width=3in]{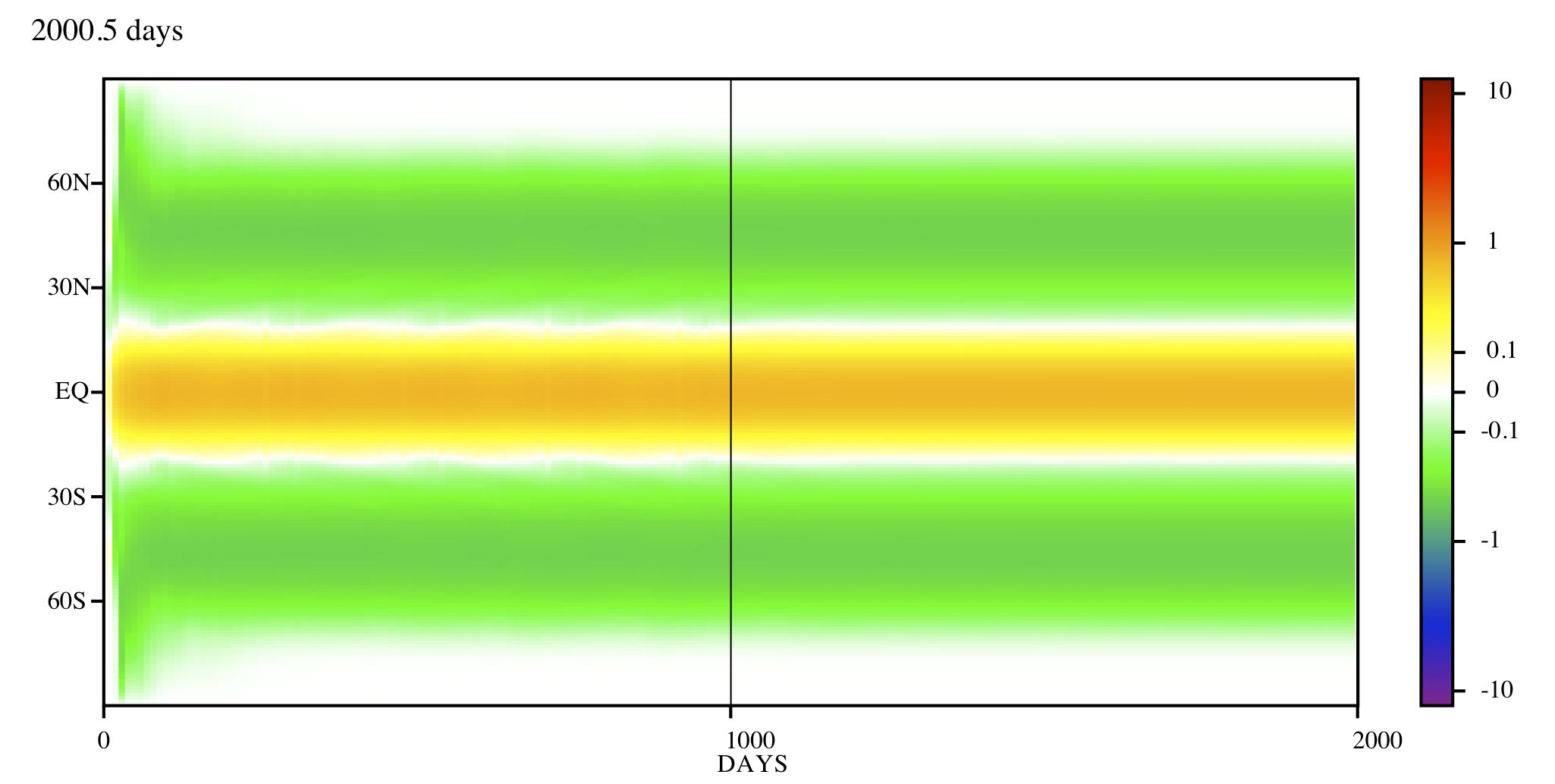}
\caption{Timelines of the zonal mean relative vorticity (left) and zonal velocity (right) as calculated for pure hydrodynamic problem with no imposed toroidal field ($B_0 = 0$).  Top: DNS on a spherical geodesic grid with $D = 163,842$ cells.  Bottom: DSS (CE2) with $L = 100$ (bottom).  A running time average commences at the midpoint in time (vertical line) at which point statistics are accumulated.}
\label{figure2}
\end{figure}

The history and statistics of these hydrodynamic jets for DNS is
displayed in the timelines in the upper panels of
Figure~\ref{figure2}. In the left portion of each panel the relative
vorticity and relative zonal velocity (averaged over a period of 10
days) is shown as a function of latitude and time. At $1000$~days
(half-way through the evolution --- signified by a vertical line in
the figures) temporal averaging is switched on and a running average
from that point is displayed in the figures. This running average
eventually settles down to show the mean position and strength of the
jets. 

\begin{figure}
\includegraphics[width=6in]{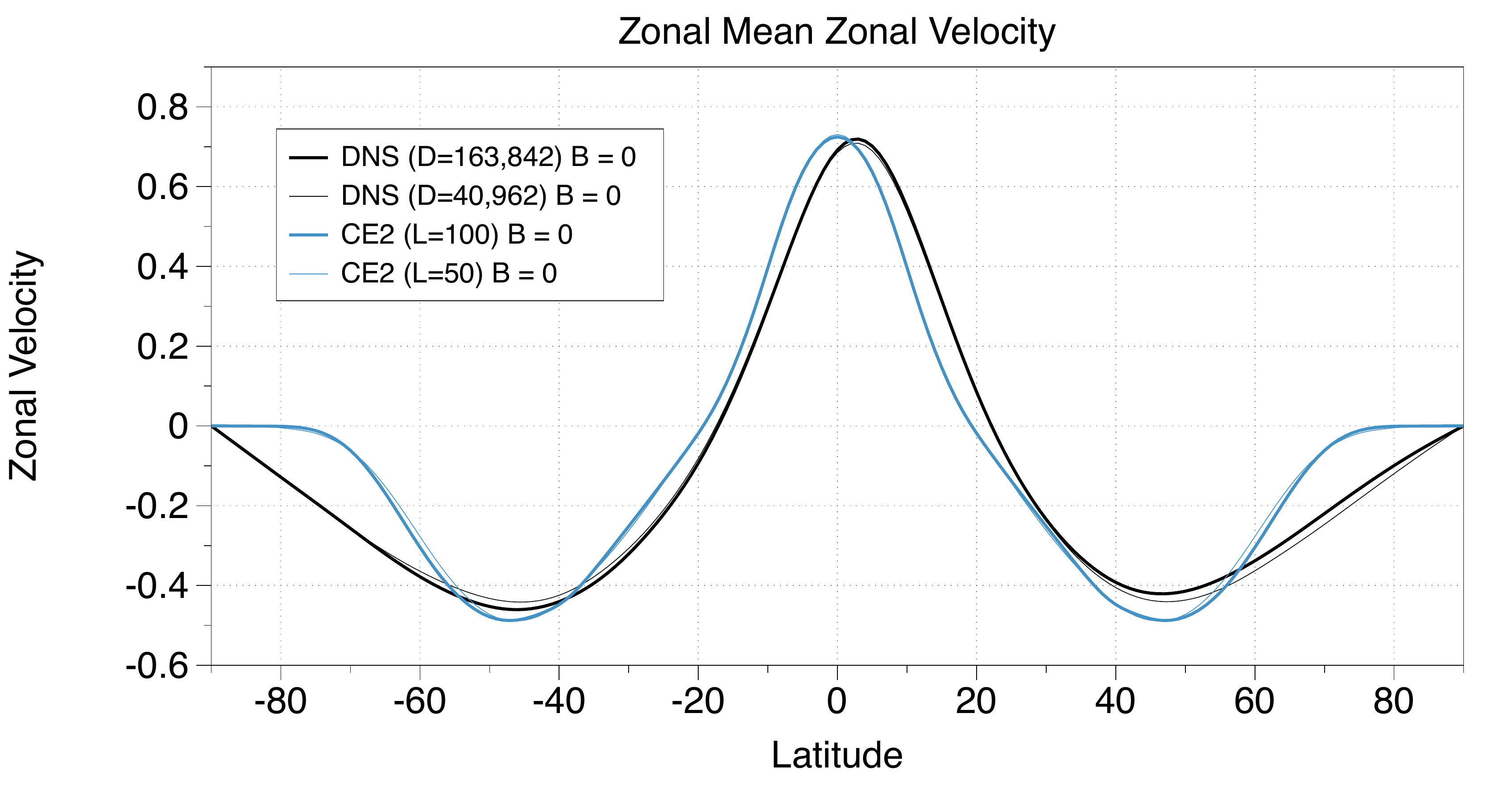}
\caption{Comparison of the mean zonal velocity as calculated in DNS and DSS (CE2) in the pure hydrodynamic problem with no imposed magnetic field ($B_0 = 0$).  Convergence with increasing resolution is evident both for DNS and for DSS (CE2).  The prograde jet is reproduced well by CE2.  Due to the finite time interval of accumulating statistics (1000 days), statistics obtained from DNS are not perfectly symmetric about the equator.}
\label{figure3}
\end{figure}

Figure~\ref{figure2} compares these timelines with those calculated by DSS for the same parameter values. The direct statistical simulation achieves remarkable agreement with the DNS in both the position and strength of the jets. This is confirmed in Figure~\ref{figure3} which demonstrates that the time-averaged zonal mean zonal velocity as calculated by DSS agrees well with DNS except at high latitudes.  Moreover, whilst the DSS respects the  north-south symmetry as expected, for the DNS the average position of the prograde jet is slightly off-equator, reflecting the finite length of data over which the averages are calculated.
Figure~\ref{figure3} also demonstrates that good convergence with
increasing resolution is achieved, both for DNS and for DSS.  

That DSS and DNS agree well is reflected in the data recorded in
Table~1. There we give some non-dimensional ratios that can only be
calculated once the kinetic and magnetic energy are in a statistically
steady state.\footnote{We note that sufficient averaging must be
  employed in order to obtain meaningful averages in both DSS and
  DNS. This is easy to achieve for DSS but is problematic for DNS.} 
These are defined as 
\begin{equation}
Ro = \frac{\langle u^2+v^2 \rangle^{\frac{1}{2}}}{2 \Omega L}, \quad
Rm = \frac{\langle u^2+v^2 \rangle^{\frac{1}{2}}L}{\eta},
\end{equation}
where we recall that $\Omega=2 \pi$, $L=1$ and $\eta=10^{-4}$.
That DSS is able to reproduce the jets using a cumulant hierarchy
truncated at second order is an interesting result. It is evidence
that the forward enstrophy cascade and anisotropic backward energy
cascade \citep{Kraichnan:1980p227,salmon98}, which is frequently
invoked to explain their existence, is in fact not necessary ---
there can be no cascade in the absence of eddy-eddy interactions.  We
note here that the enstrophy cascade argument has also been questioned
in the context of planetary atmospheres
\citep{Vallis1992,Schneider06a,OGorman:2007p76} where shearing and
modification of the thermal structure of the atmosphere by eddy fluxes
weaken the eddy-eddy interactions.  Here it is therefore  Reynolds
stresses that are primarily responsible for the build-up of the zonal
flows.  This result is important for our understanding of the driving
of zonal flows in planetary atmospheres.

\begin{figure}
\centerline{\includegraphics[width=4in]{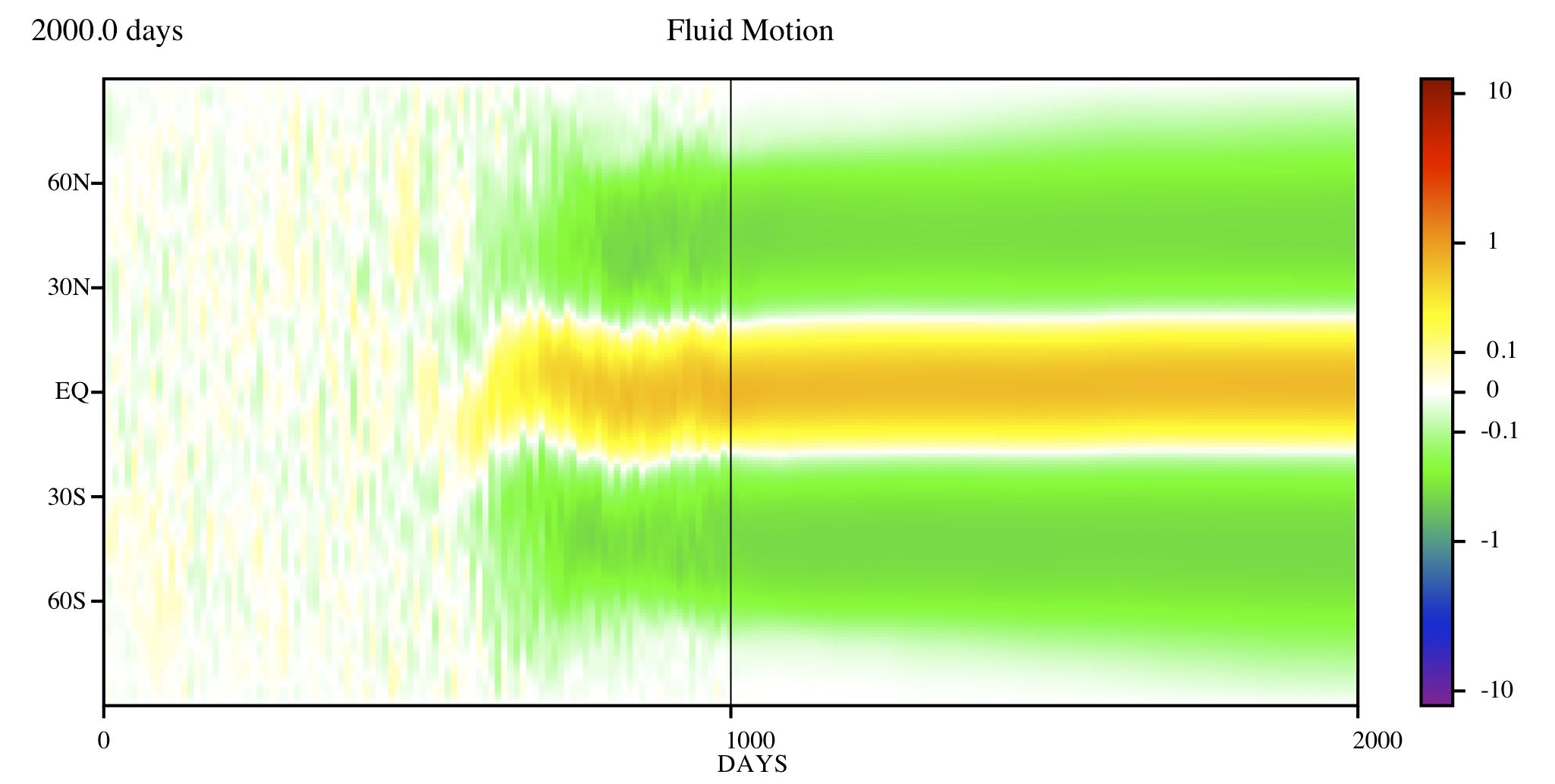}}

\centerline{\includegraphics[width=4in]{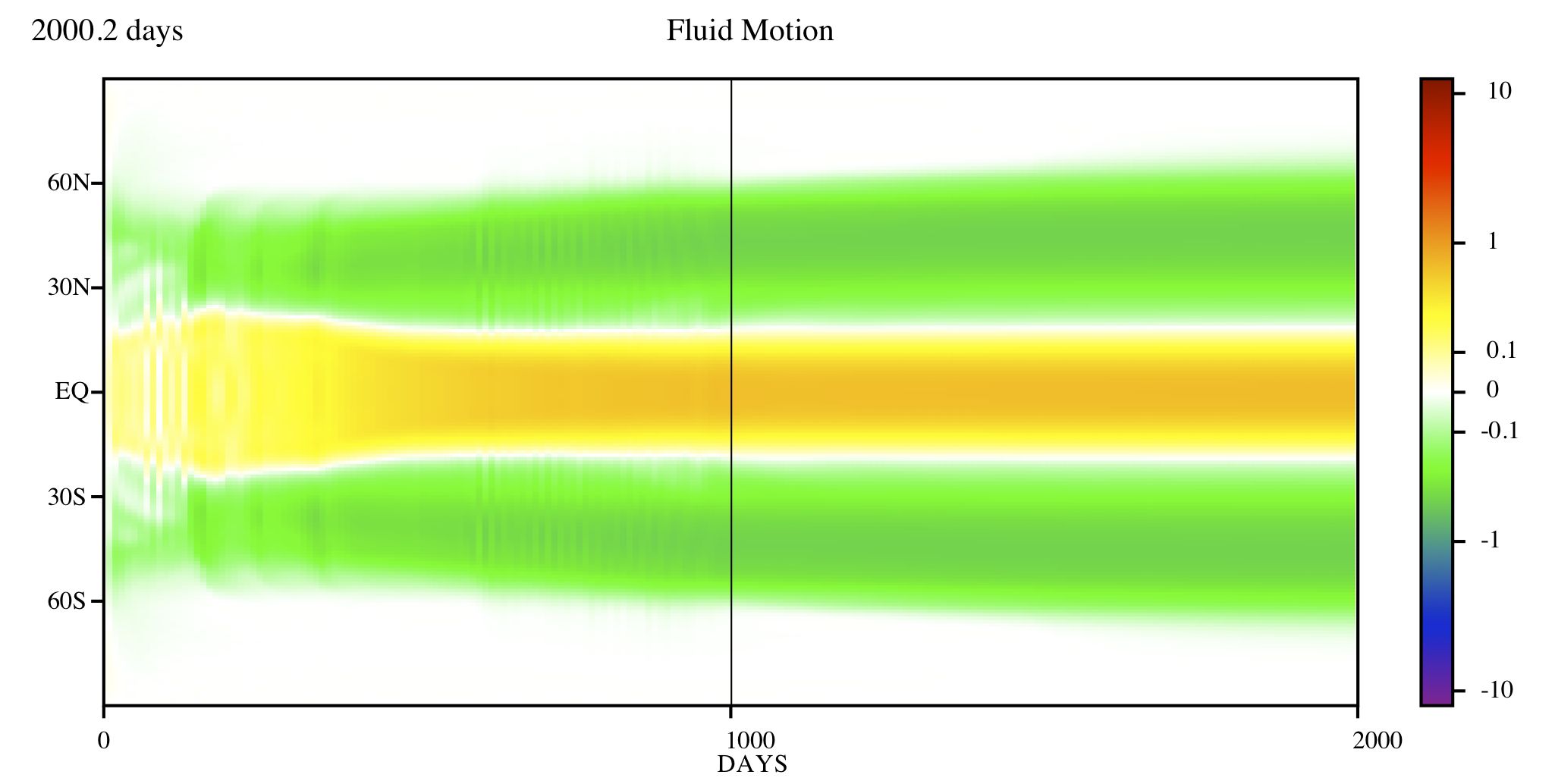}}
\caption{Timelines of the zonal mean zonal velocity for an imposed toroidal magnetic field with $B_0 = 0.1$.  Top:  DNS with $D = 40,962$ cells.  Bottom: DSS (CE2) with $L=100$.}
\label{figure4}
\end{figure}

We now examine the effects of including a toroidal magnetic field and examine the dynamics for two imposed field strengths, $B_0$.
For $B_0 = 0.1$ the onset of jet formation  is delayed but in both DNS
and DSS the system eventually settles into a statistically steady
state as shown in Figure~\ref{figure4}. For both methods, for this
relatively weak imposed mean field, there is eventually little
suppression of the jets, with slightly more suppression occurring in
the DSS. 
For this choice of parameters, the magnetic energy is small compared
with the kinetic energy of the flow (4\% for both DNS and DSS (CE2)),
and so it is to be expected that the role of the magnetic field will
be secondary. Moreover the magnetic field has been expelled to high
latitudes by the strong jets and turbulence at low latitudes (see
Figure~\ref{figure7}). This flux expulsion
\citep{Weiss:1966p605,Tao:1998p608} leads to separated regions with
different dynamics; at low latitudes, where the field is weak, the
hydrodynamic evolution continues unimpeded, whilst at high latitudes
the magnetic field leads to some suppression of the jets. 

\begin{figure}
\includegraphics[width=3in]{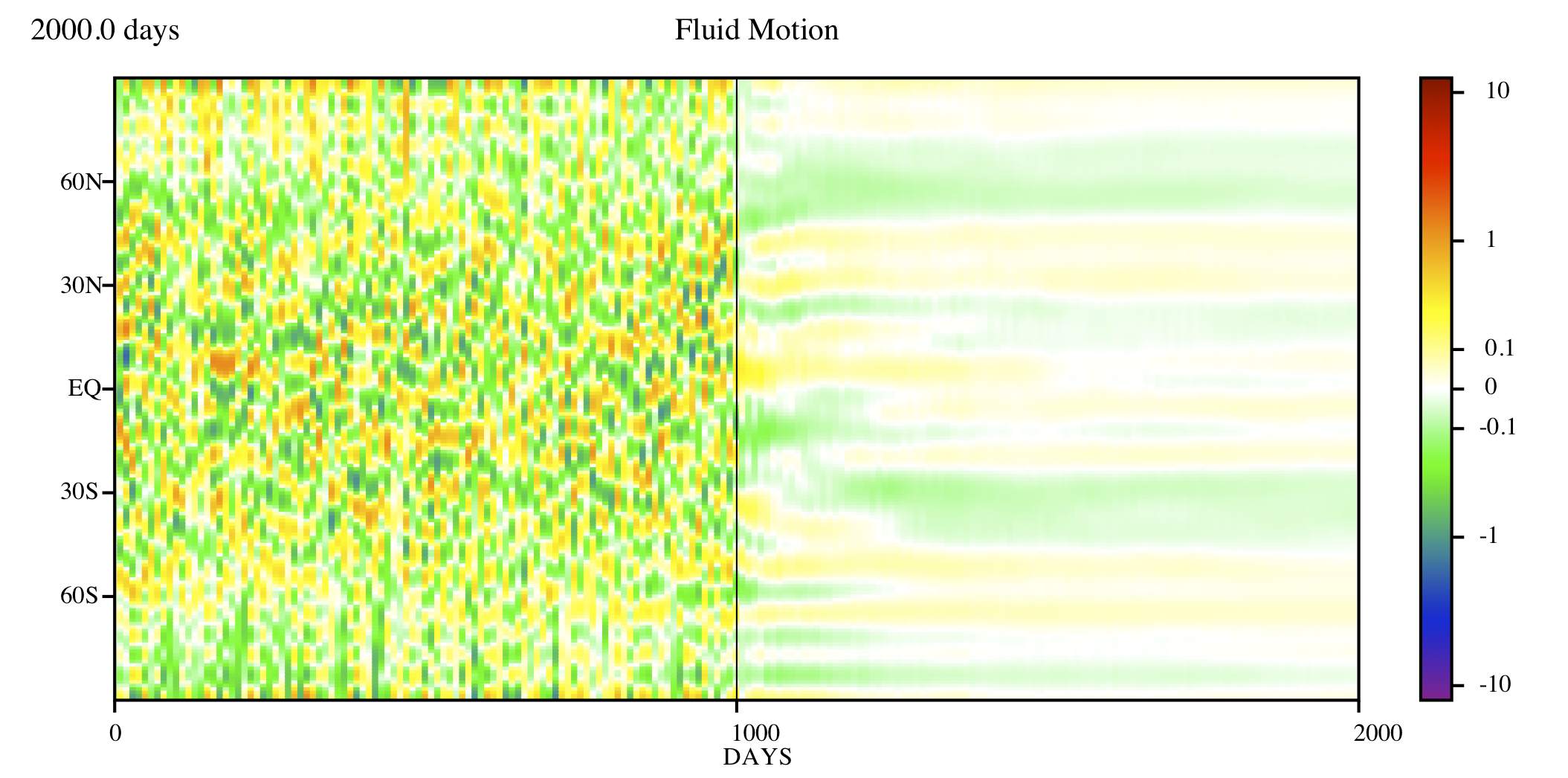}
\includegraphics[width=3in]{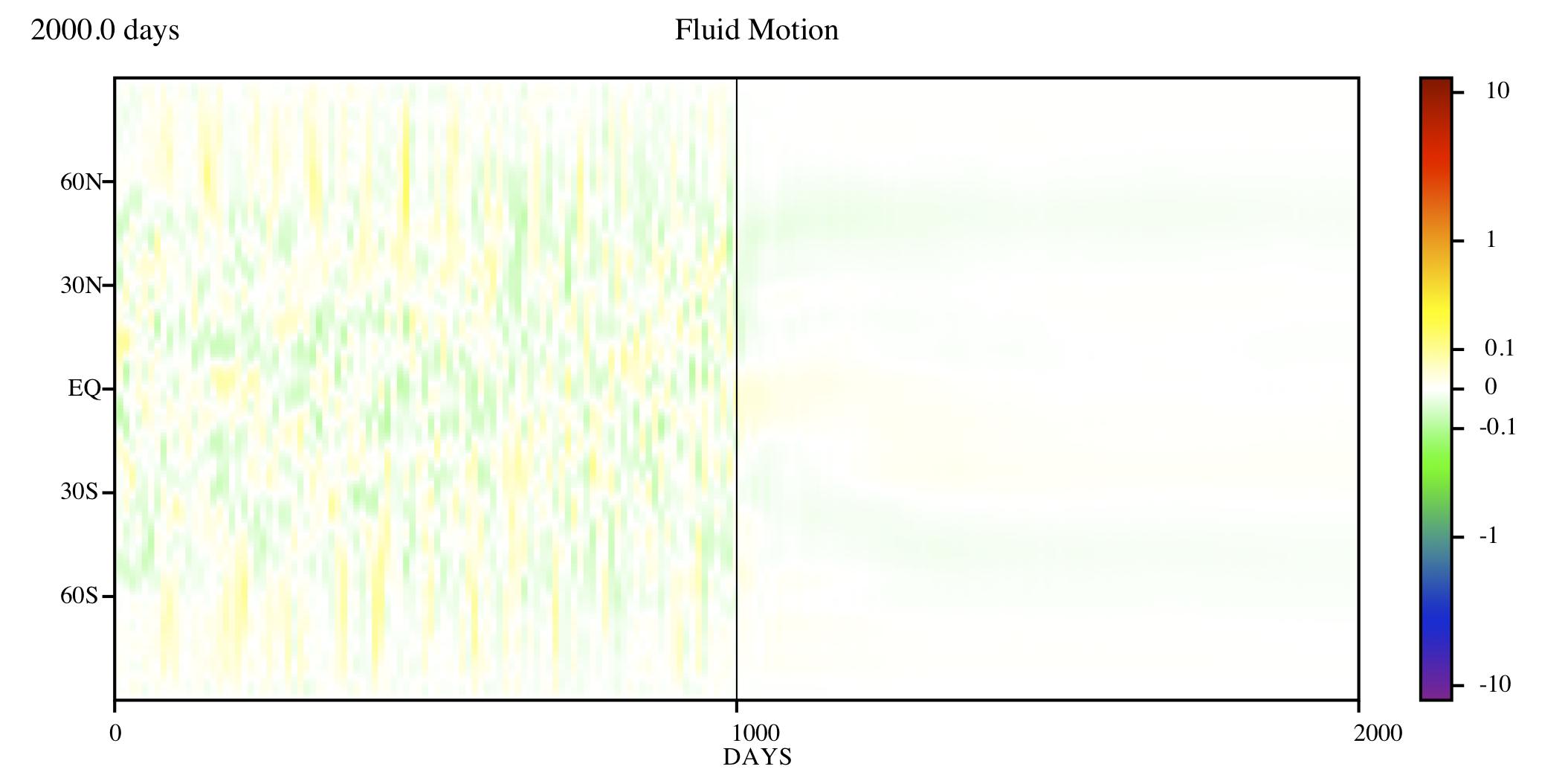}

\includegraphics[width=3in]{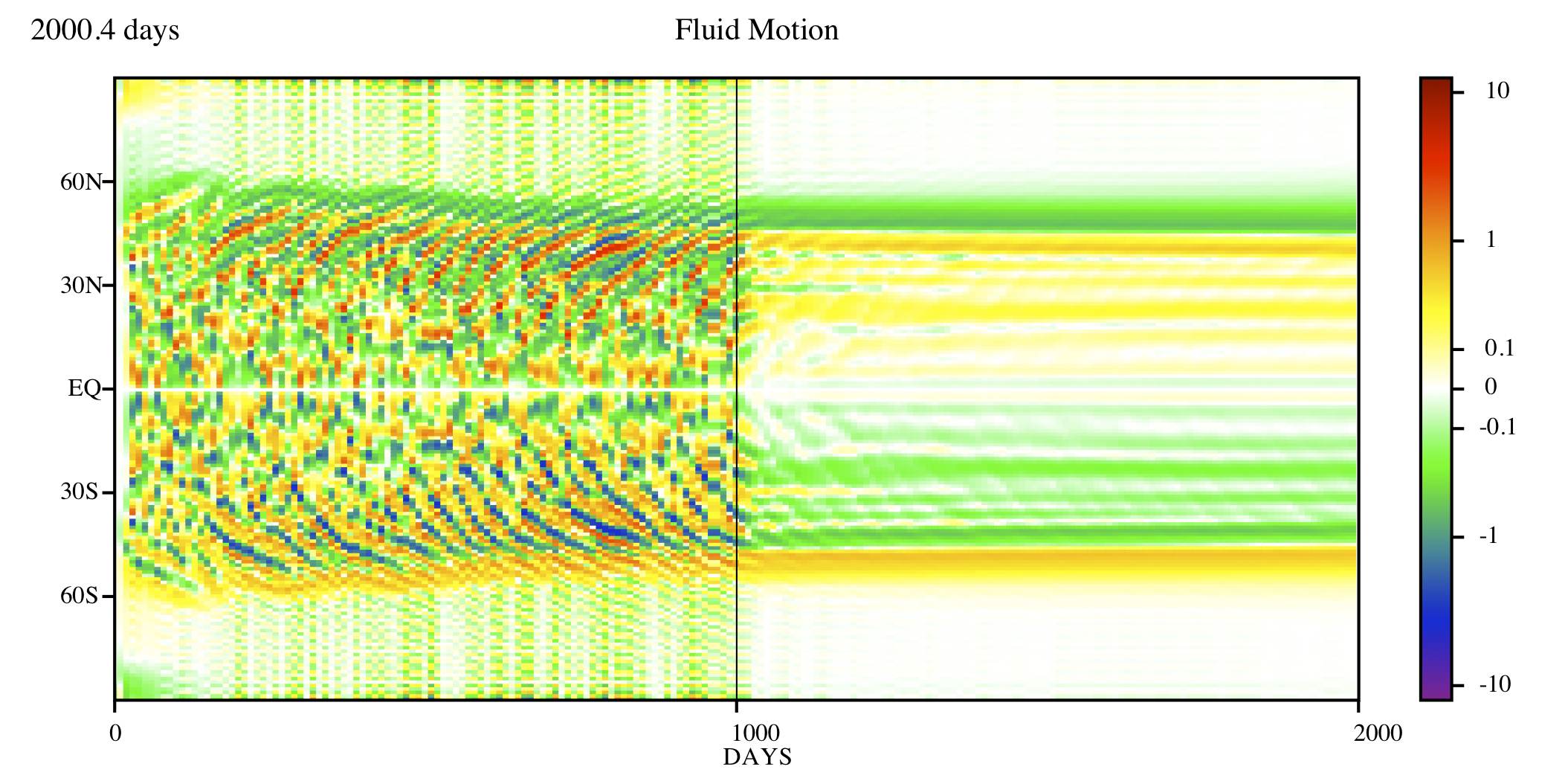}
\includegraphics[width=3in]{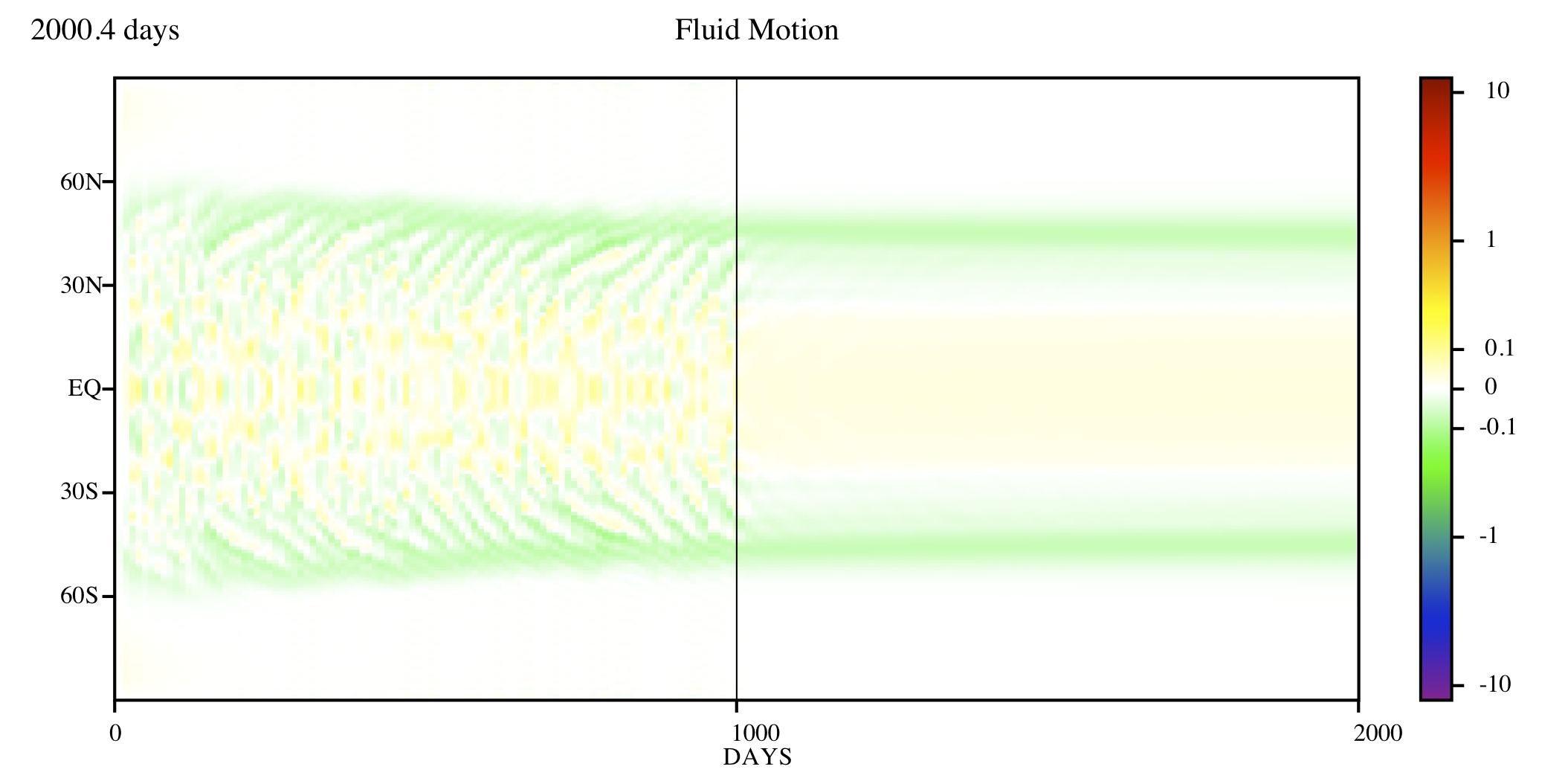}
\caption{Same as Fig. \ref{figure2} except with $B_0 = 0.5$.}
\label{figure5}
\end{figure}

\begin{figure}
\includegraphics[width=6in]{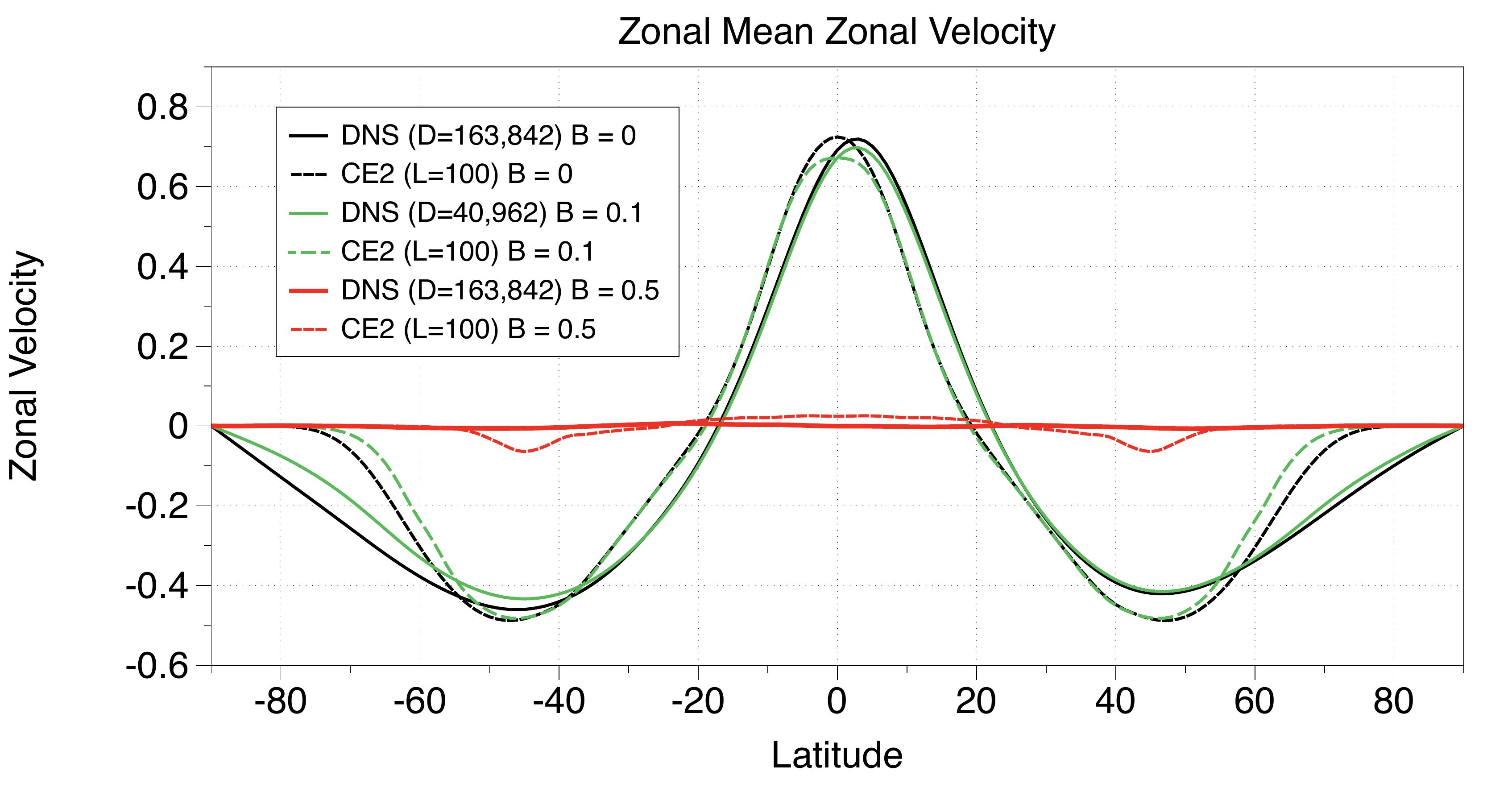}
\caption{Comparison of the mean zonal velocity as calculated in DNS and DSS (CE2) for imposed toroidal fields of $B_0 = 0$, $0.1$ and $0.5$.}
\label{figure6}
\end{figure}

\begin{figure}
\includegraphics[width=6in]{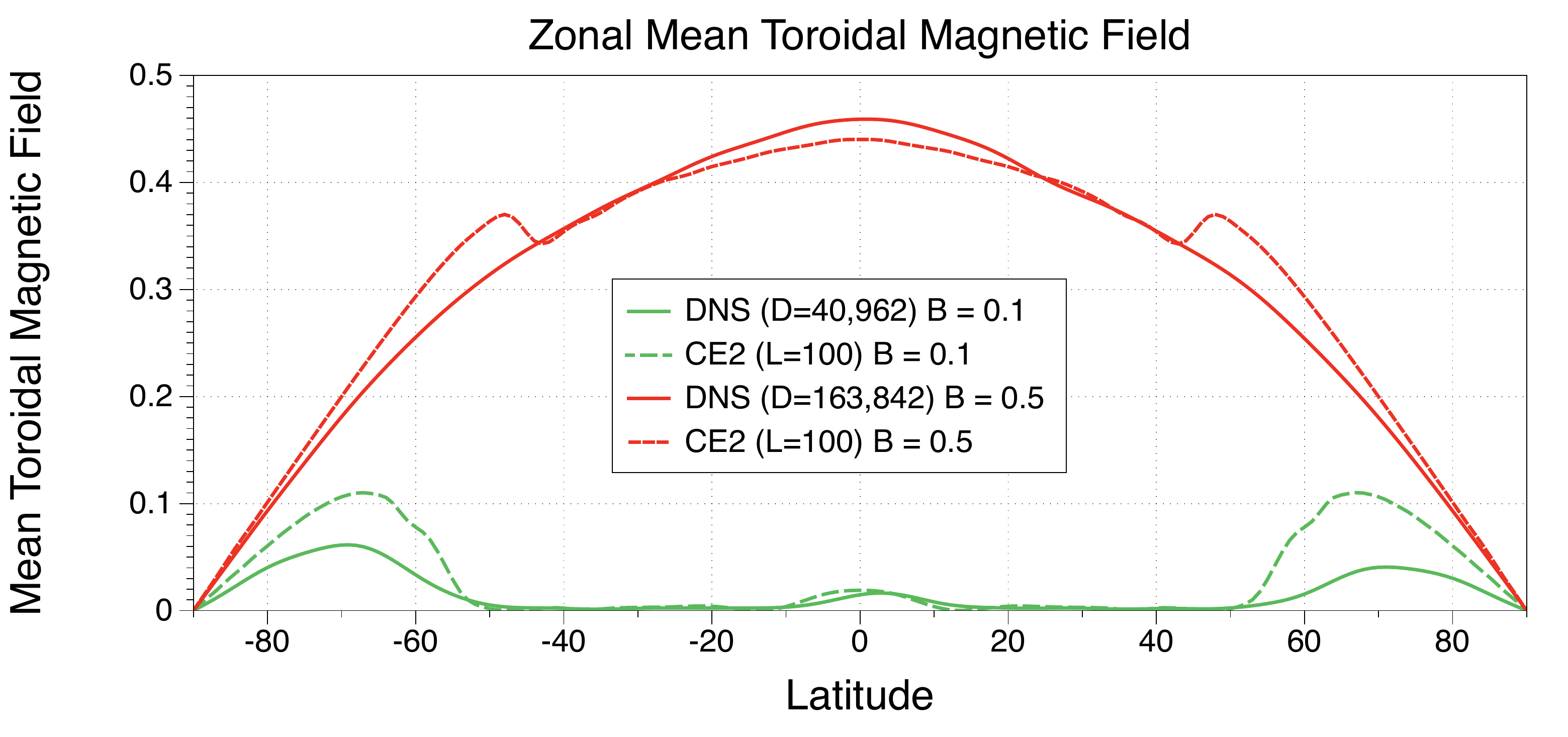}
\caption{Comparison of the mean toroidal magnetic field as calculated in DNS and DSS (CE2) for imposed toroidal fields of $B_0 = 0.1$ and $0.5$.}
\label{figure7}
\end{figure}

At $B_0 = 0.5$, however, strong qualitative changes are plainly evident as the jets are destroyed by the fluctuations in the magnetic field, both in DNS and in DSS, in agreement with the findings of \cite{Tobias:2007p116}.  Small remnants of the jets persist in DSS at high latitudes, where the imposed toroidal field is weakest (see Fig. \ref{figure5}).  DNS results (top panels) show the incoherent nature of the flows ---  it is this that leads to the suppression of the jets. For this case, the magnetic energy is in approximate equipartition with the kinetic energy, as shown in Table~1. Once again DNS and DSS show a remarkable agreement for this case; see Fig. \ref{figure6}.  A comparison of the mean toroidal component of the magnetic field is shown in Fig. \ref{figure7}.  Here the situation is reversed from the previous weaker field case. The magnetic field here is too strong to be expelled by the eddies and the jet never forms at low latitudes. Therefore the field is confined to low latitudes and the (weaker) jets can only be found at high latitudes where the imposed field is weaker. For this strength of imposed field the kinetic energy is reduced (see the values of $Rm$ in Table~1) and the magnetic energy comes into equipartition with the kinetic energy.
Clearly the transport of angular momentum by the Reynolds and Maxwell stresses and of magnetic flux by the turbulent advection has acted in a very different manner here. The discussion of these processes and their description via the second cumulants is postponed to a subsequent paper.

\subsection{Case $L_{min} = 8$, $M_{min} = 1$, and $L_{max} = 10$}

\begin{figure}
\includegraphics[width=3in]{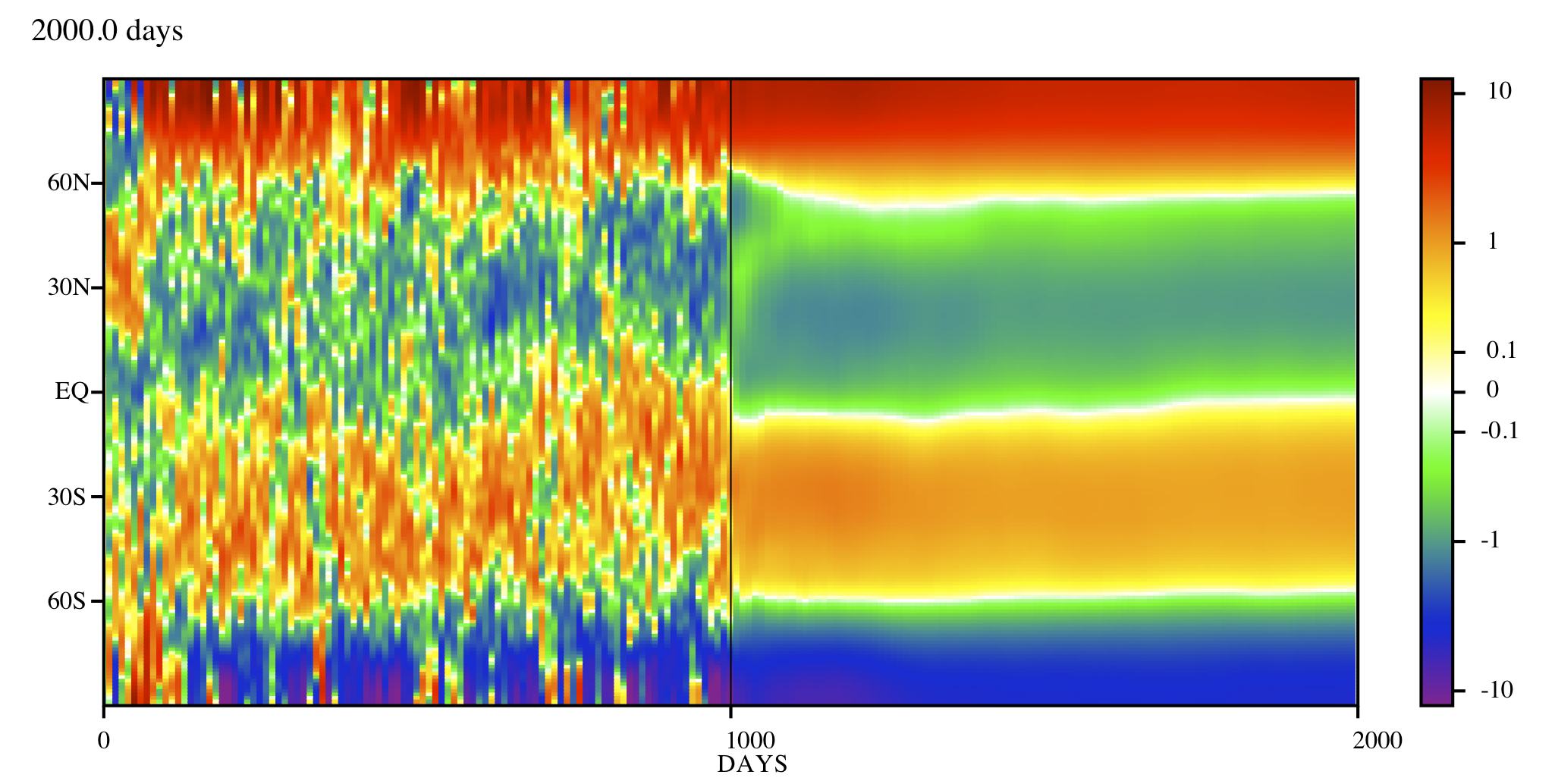}
\includegraphics[width=3in]{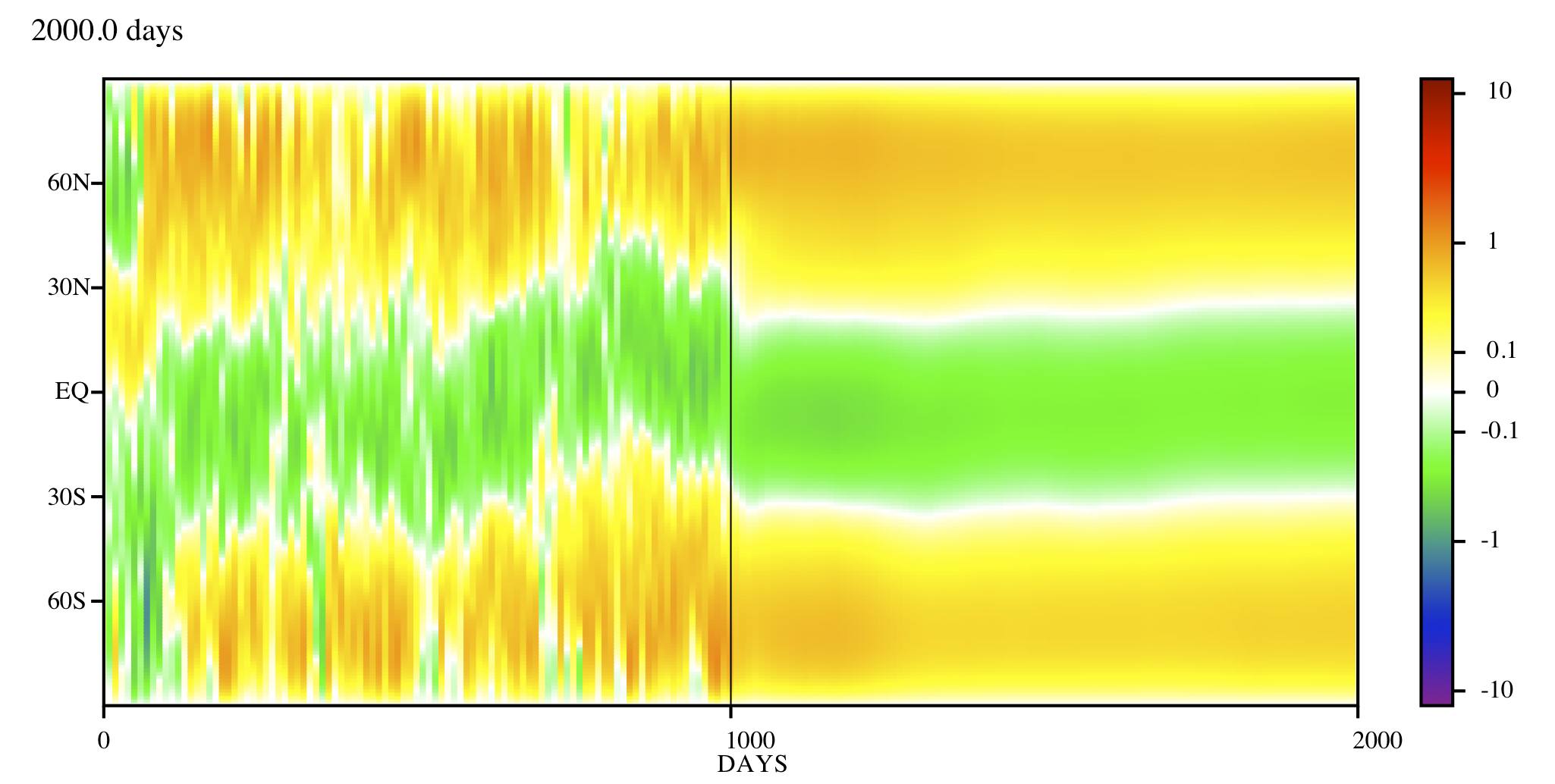}

\includegraphics[width=3in]{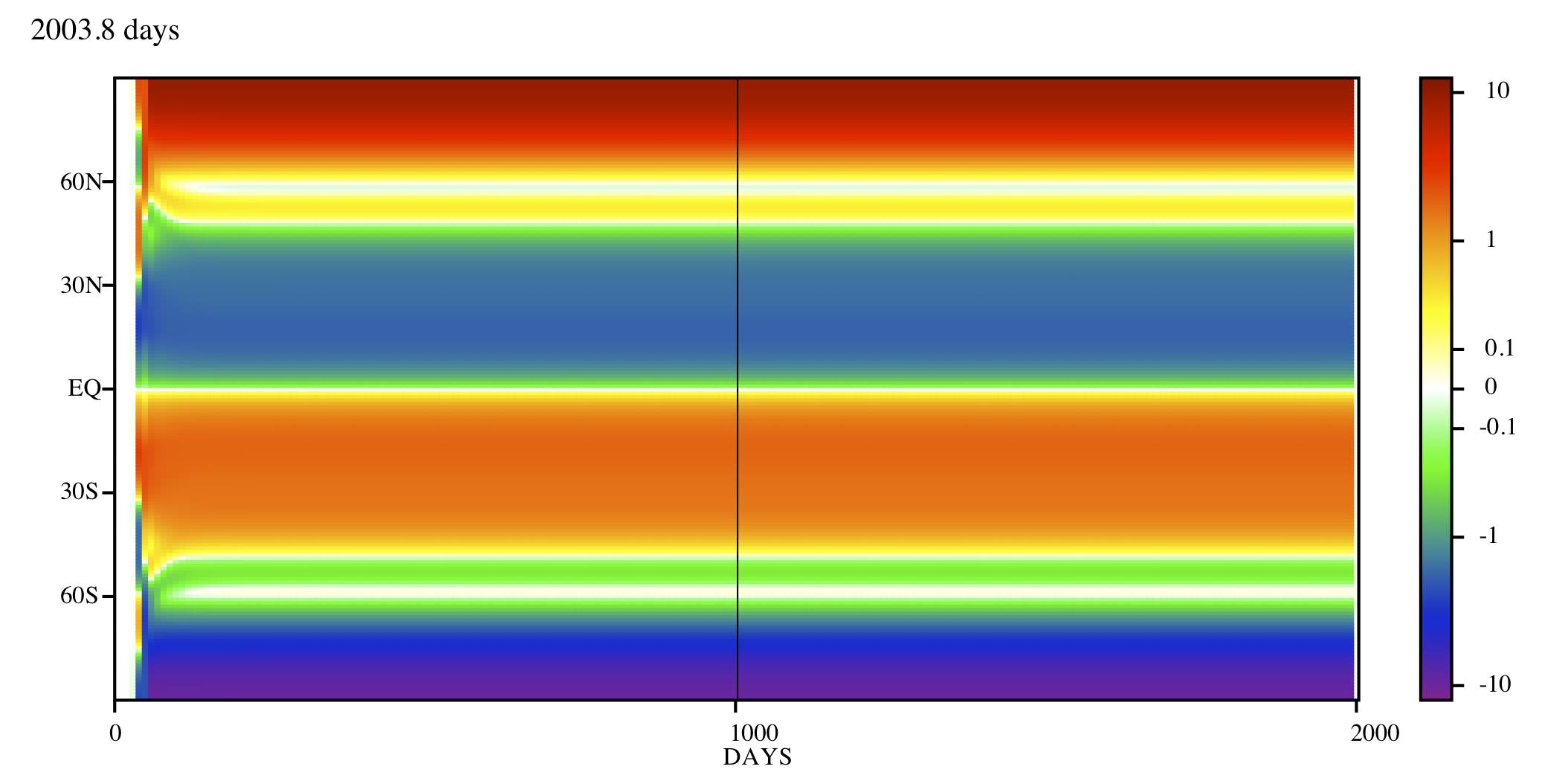}
\includegraphics[width=3in]{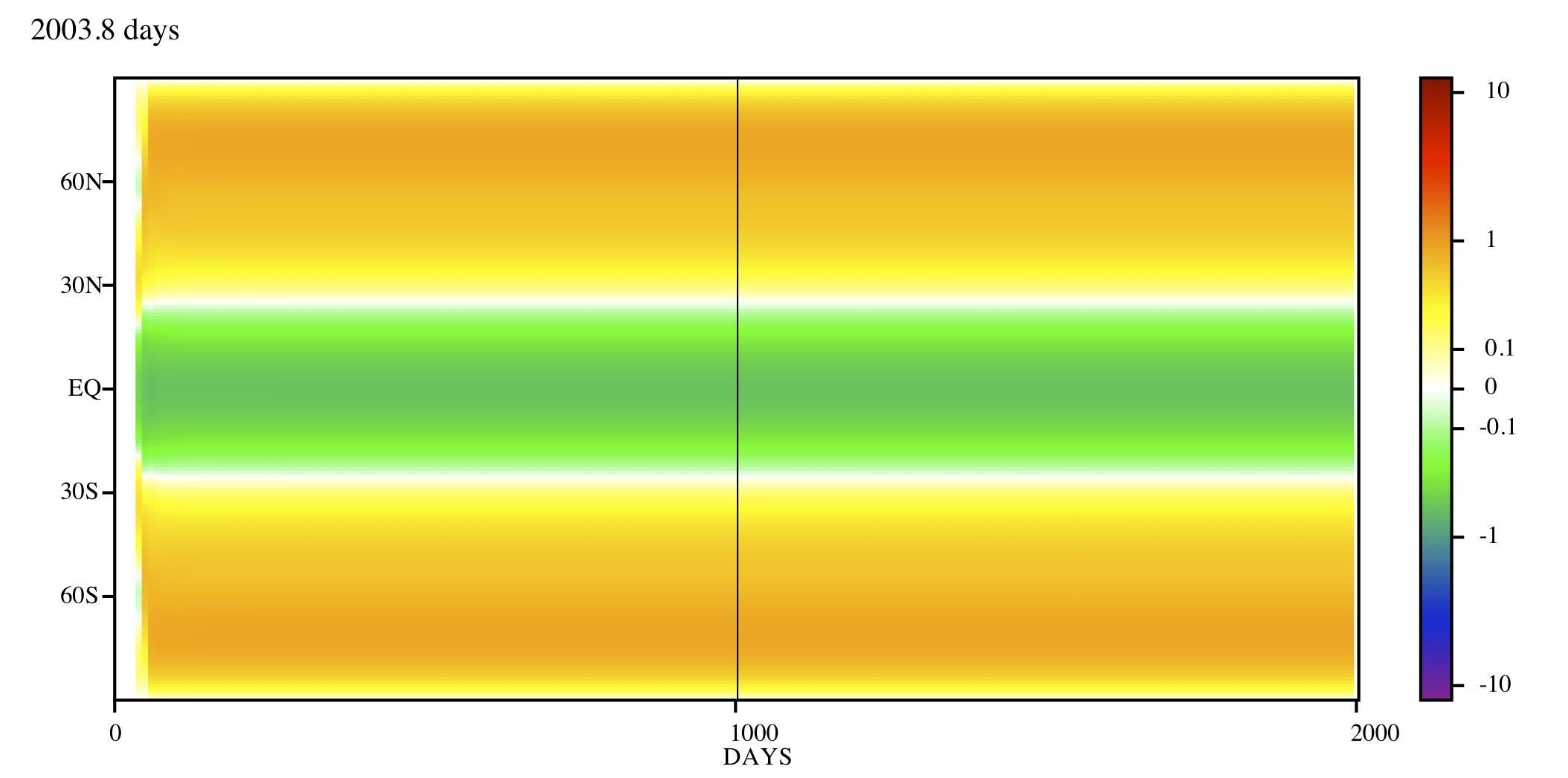}
\caption{Timelines of the zonal mean relative vorticity (left) and zonal velocity (right) as calculated for pure hydrodynamic problem with no imposed toroidal field ($B_0 = 0$).  Top: DNS on a spherical geodesic grid with $D = 163,842$ cells.  Bottom: DSS (CE2) with $L = 100$ (bottom).}
\label{figure8}
\end{figure}

\begin{figure}
\includegraphics[width=6in]{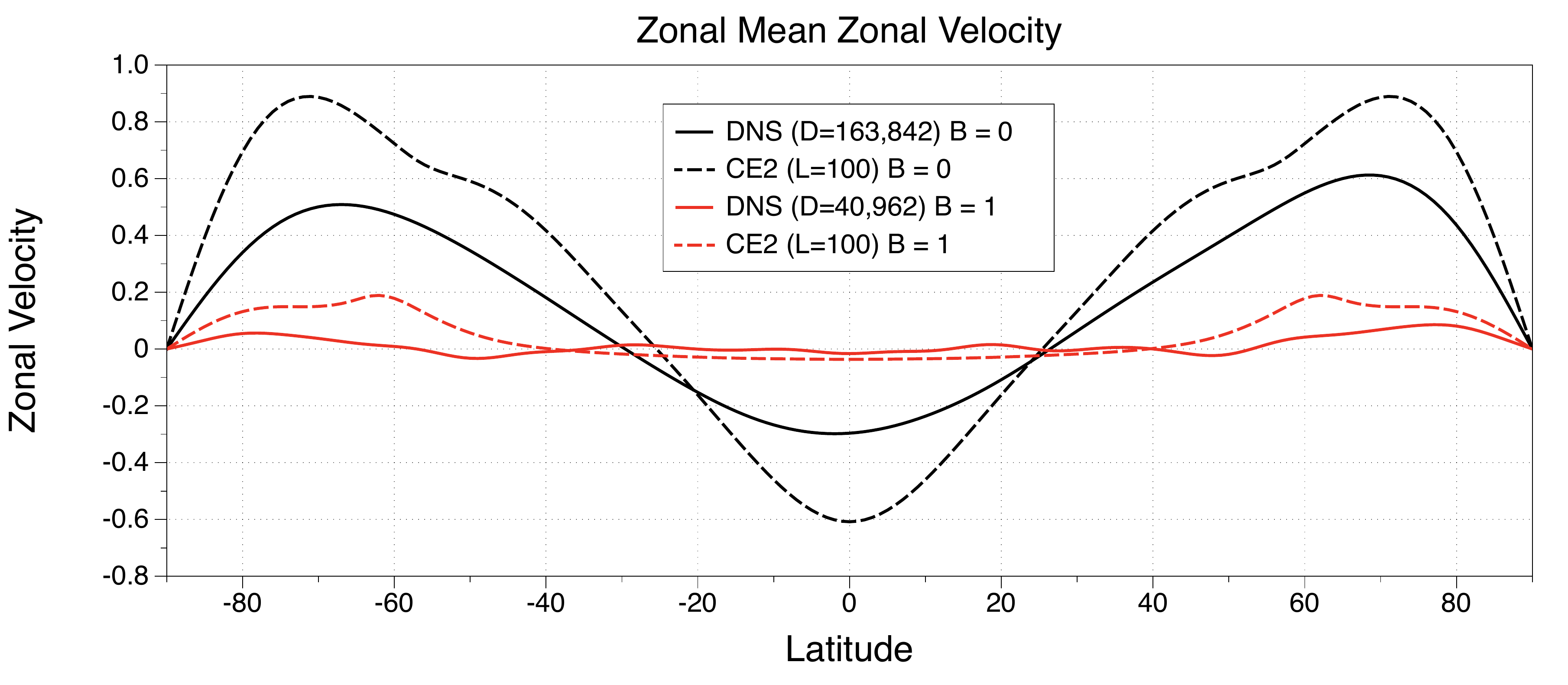}
\caption{Comparison of the mean zonal velocity as calculated in DNS and DSS (CE2) for imposed toroidal fields of $B_0 = 0$ and $1.0$.}
\label{figure9}
\end{figure}

We now consider the effect of reducing the minimum stochastic forcing wavevector in the azimuthal direction, $M_{min}$, down to wavenumber 1.  This brings stochastic effects to larger scales and so presents a more robust challenge for DSS.   
A comparison of the zonal mean relative vorticity and zonal velocity as calculated by DNS and DSS (CE2) is shown in Fig. \ref{figure8} for the hydrodynamic problem.  In contrast to the previous case, there are two prograde jets at high latitudes, and one equatorial retrograde jet; again the total angular momentum is close to zero.  Now, however, the jets are seen to wander significantly in latitude in DNS owing to the continual random forcing at large zonal scales. Once established in DSS, however,  they remain fixed in place.  As a consequence, the time-averaged zonal means are reduced in magnitude in DNS compared with DSS, as made apparent in the quantitative plot of Fig. \ref{figure9}.  

\begin{figure}
\includegraphics[width=3in]{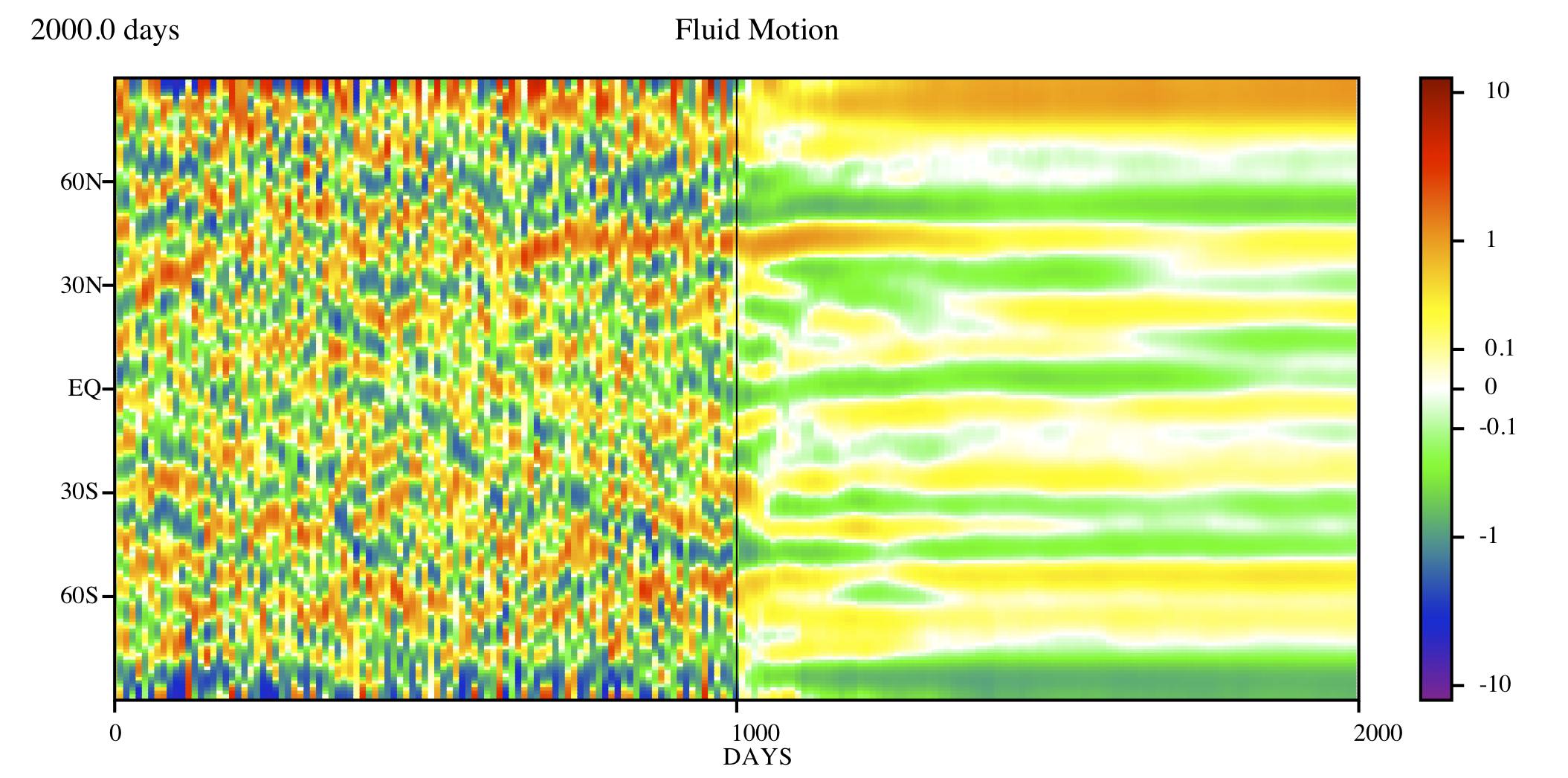}
\includegraphics[width=3in]{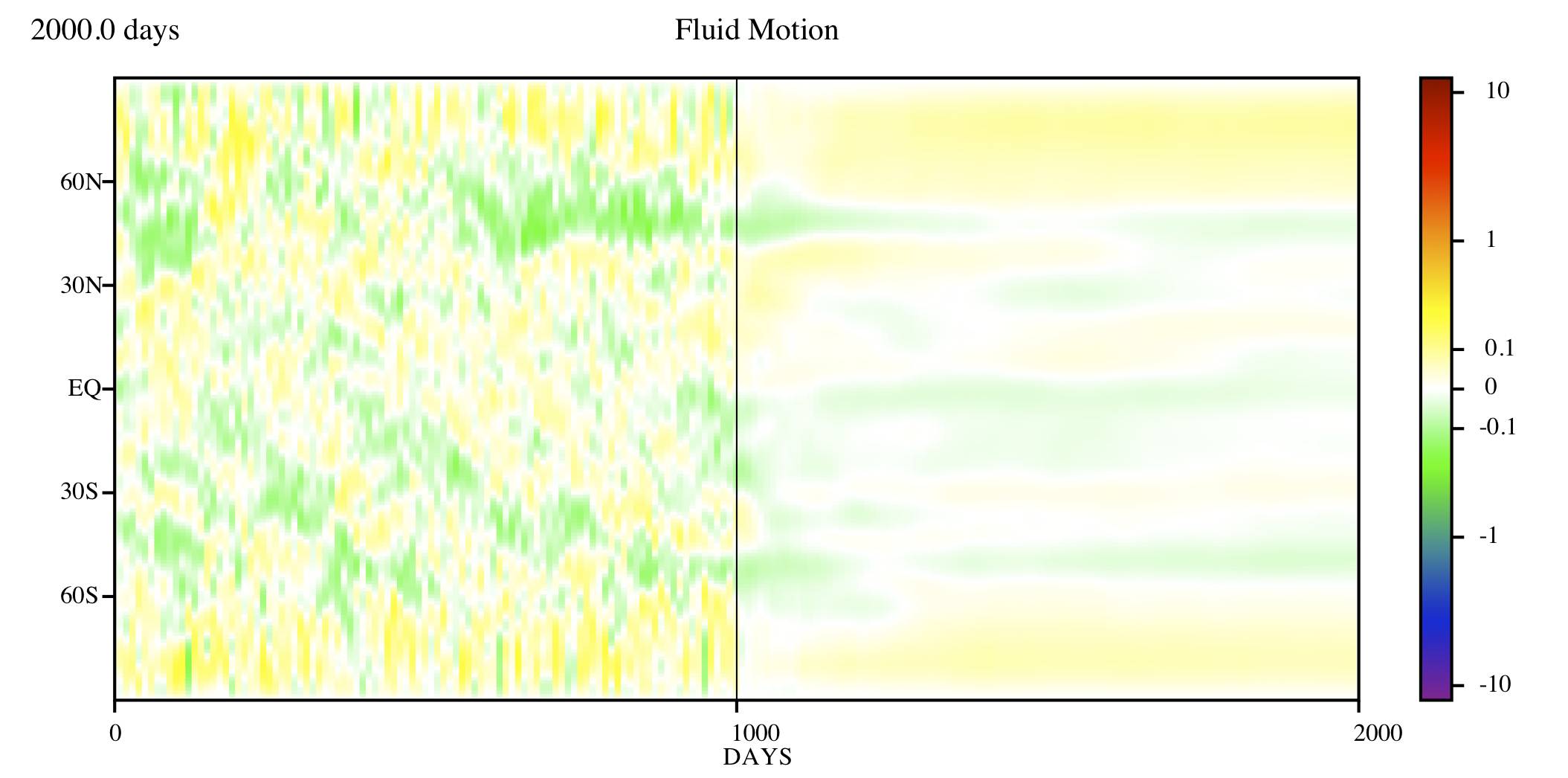}

\includegraphics[width=3in]{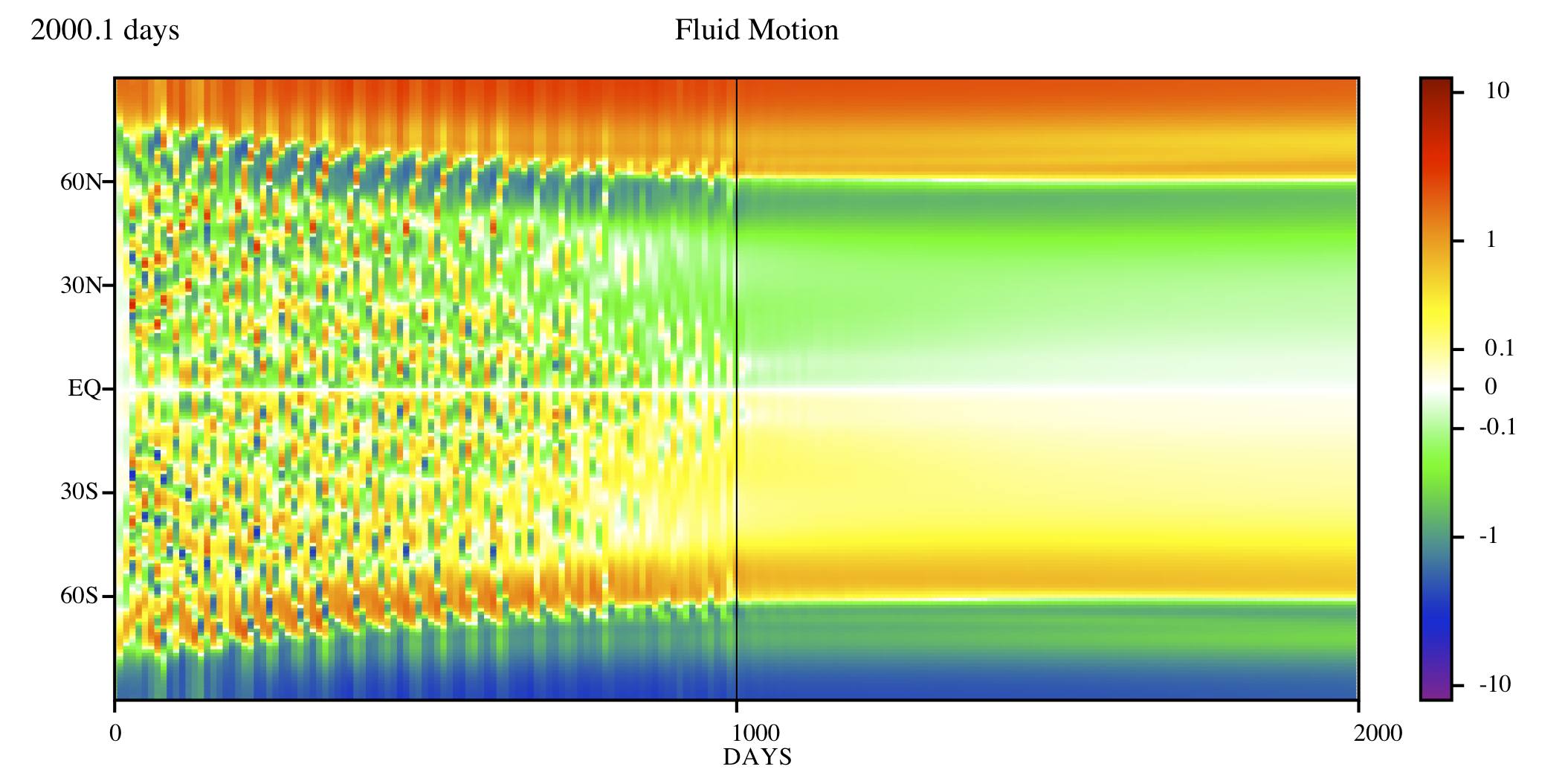}
\includegraphics[width=3in]{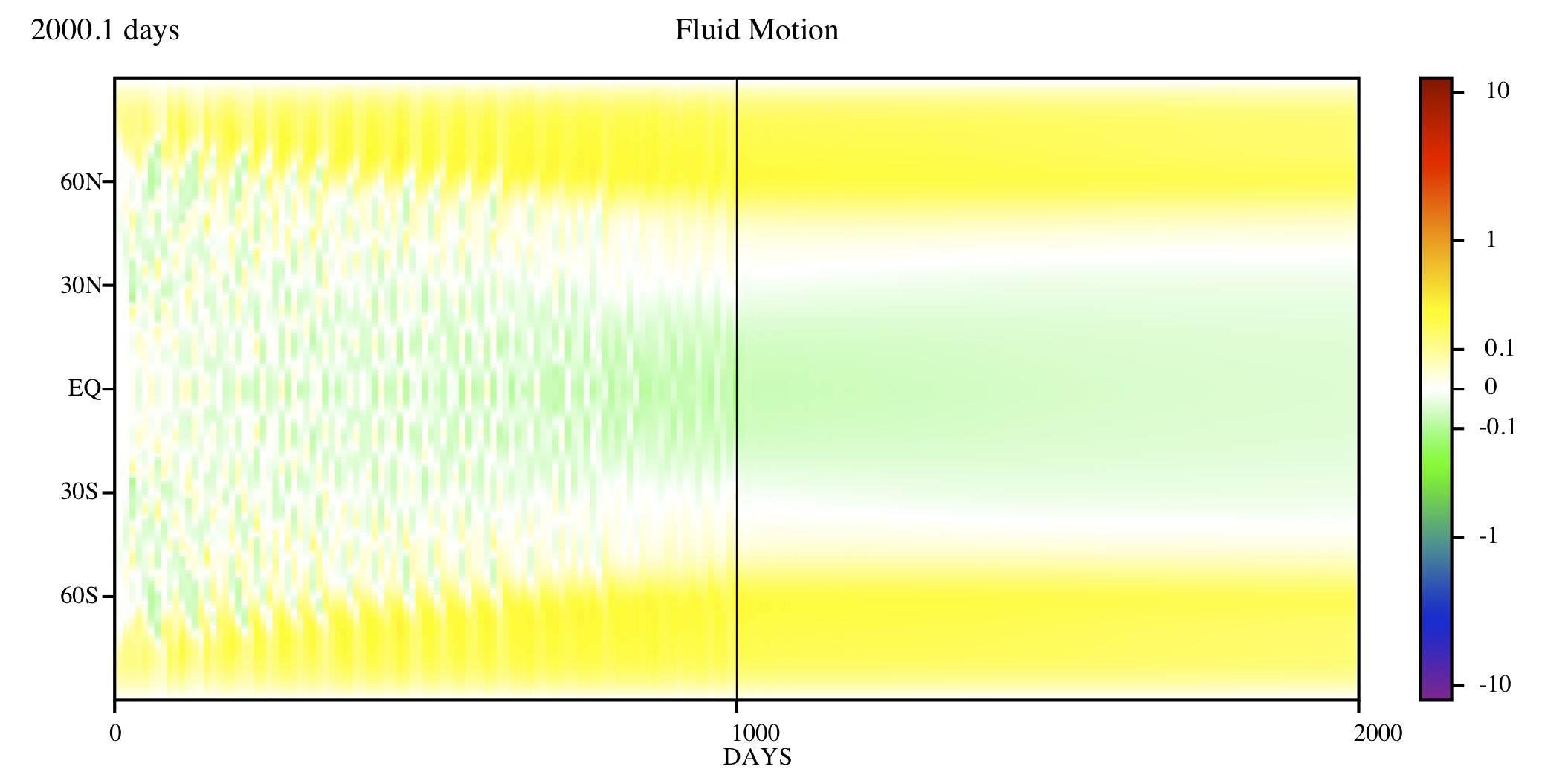}
\caption{Same as Fig. \ref{figure8} except for $B_0 = 1$ and with DNS run on a lower resolution spherical geodesic grid with $D = 40,962$ cells.}
\label{figure10}
\end{figure}

Imposing a relatively weak
toroidal field by setting $B_0 = 0.1$ again has little effect on the zonal mean velocity as shown in Figure~\ref{figure9}.  However at $B_0 = 1$ the jets are largely eliminated by the fluctuations in the magnetic field, both in DNS and in DSS.  Somewhat larger jet remnants remain at high latitudes, where the imposed toroidal field is weakest, and again stronger jets are found in DSS than in DNS (see Fig. \ref{figure10}).  

\begin{figure}
\centerline{\includegraphics[width=4in]{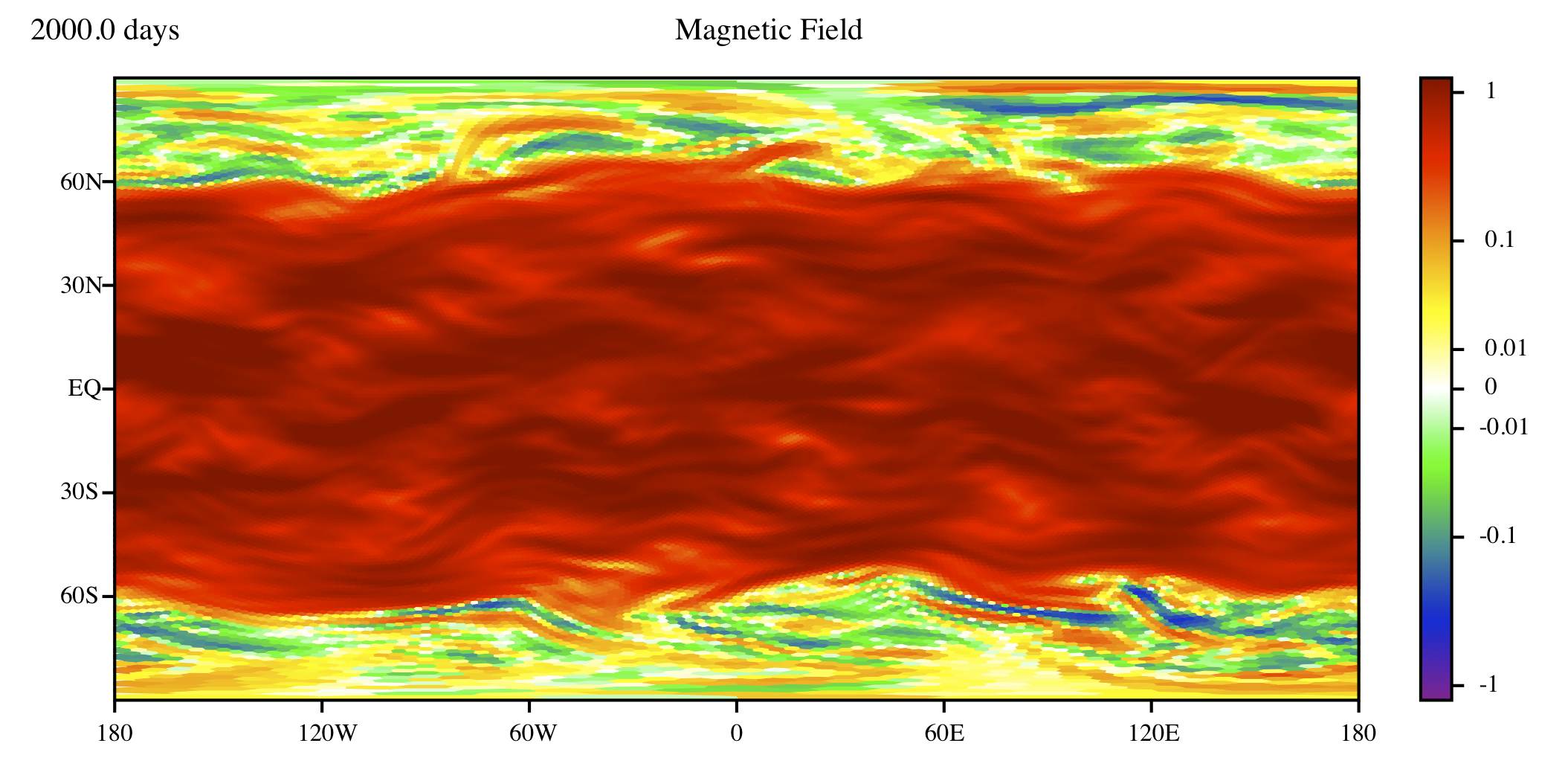}}
\centerline{\includegraphics[width=4in]{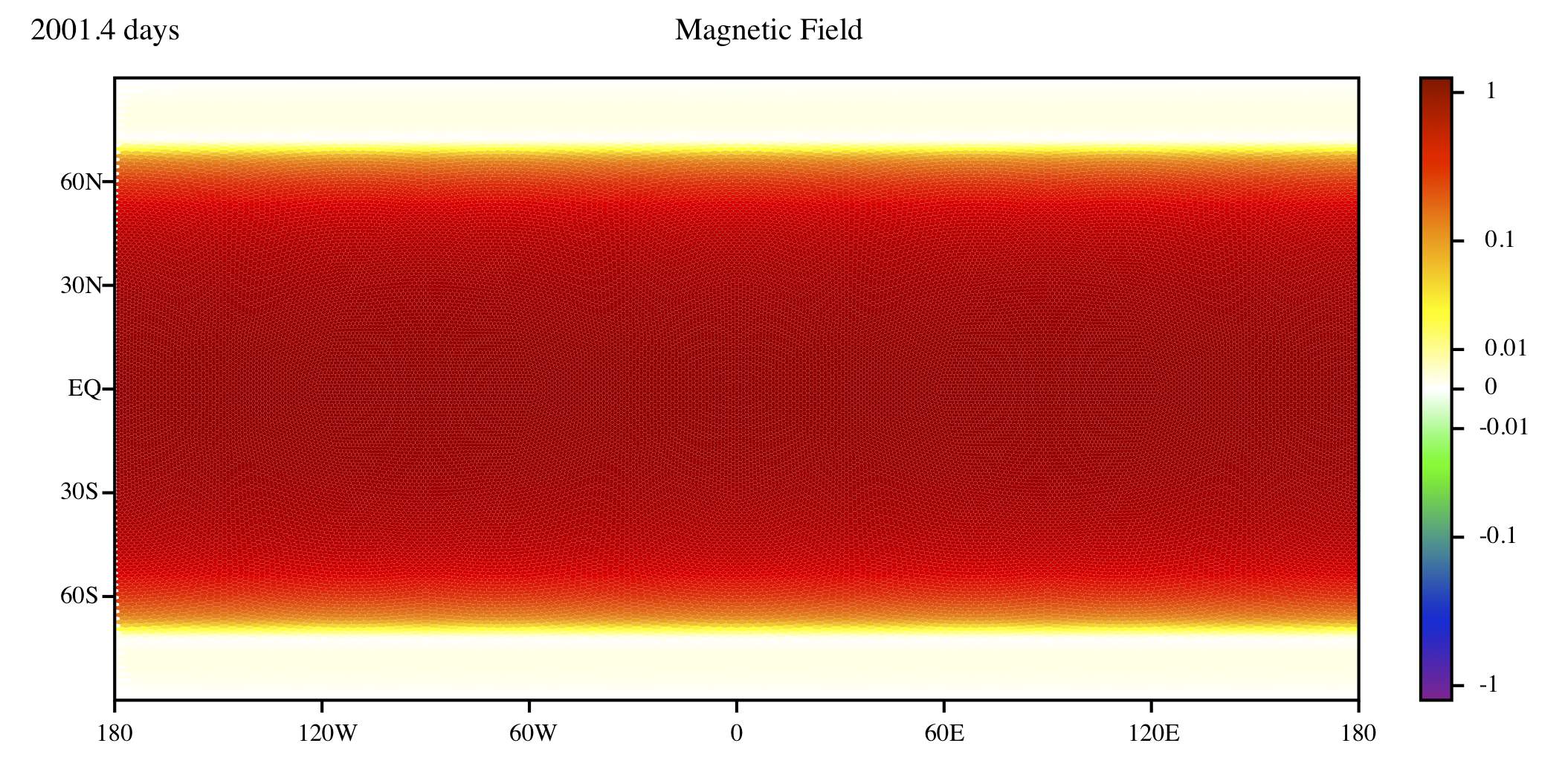}}
\caption{Toroidal component of the magnetic field for $B_0 = 1.0$.  Top:  Instantaneous field found by DNS on a spherical geodesic grid with $D = 40,962$ cells.  Bottom:  Zonal mean field found by DSS with $L = 50$.}
\label{figure11} 
\end{figure}

The toroidal field is tightly confined to latitudes less than roughly $60^\circ$, as Fig. \ref{figure11} depicts, likely owing to flux expulsion of the field.  Again DSS does a reasonably good job of reproducing the mean field.

\section{Discussion}

In this paper we have introduced the concept of Direct Statistical Simulation (DSS) for astrophysical
fluid dynamics. We have compared the results of Direct Numerical Simulation (DNS) and DSS for the problem of 
two-dimensional hydrodynamics and magnetohydrodynamics on a spherical surface. Although the set-up of the model
is relatively simple, the ensuing dynamics is not. In the hydrodynamic case, non-trivial interactions
at moderate scales drive inhomogeneous large-scale zonal flows (jets/winds). With a weak imposed field the jets remain largely unaffected and the magnetic fields are expelled to higher latitudes. With a stronger imposed toroidal magnetic field
these winds are suppressed except at high latitudes where the imposed field is weak.

We find that even the simplest formalism of DSS, based upon the truncation of the cumulant hierarchy at second order, is capable
of reproducing the driving and suppression of the zonal flows and the flux expulsion of the magnetic fields by the inhomogeneous jets.  Because the method includes interactions that are non-local in space it is very well suited to such inhomogeneous problems thatypically arise in astrophysics.
Such a truncation is equivalent to keeping the mean/eddy interactions in the eddy
equations and the eddy/eddy interactions in the mean equations, whilst suppressing the eddy/eddy interactions in the eddy equations.  Thus it is the Reynolds and Maxwell stresses that respectively drive and suppress the jet, not an inverse cascade as is frequently assumed.  

The DSS scheme is more numerically efficient than the corresponding DNS.
We believe that the results presented here are an encouraging beginning for the concept of DSS in astrophysical fluid dynamics. It is important though to determine the
range of validity of such a procedure. Clearly the method is designed to work best when the dynamics leads to the generation of substantial statistical means 
(e.g. mean flows or magnetic fields) or involves the interactions of prescribed (usually inhomogeneous) mean quantities with smaller scale turbulence.
The method is inefficient when the turbulence is dominated completely by small-scales and is largely homogeneous --- for example homogeneous, isotropic hydrodynamic turbulence, or the
small-scale dynamo problems \citep[see e.g.][]{tcb:2011}. We do believe, however, that many cases of astrophysical interest {\it do} fall into the category where DSS techniques may prove useful. 
Examples currently under consideration include the interaction of mean magnetic fields and shear flows either on a spherical surface (leading to joint instability \citep[see e.g.][]{Cally:2003p272}) or in a cylindrical
domain (leading to magnetorotational instability), the instability and mixing of large-scale shear flows in the presence of a magnetic field, and the driving of zonal flows via convection
in a tilted cylindrical annulus \citep[see e.g.][]{Brummell:1993p2}. It will be interesting to determine how well the techniques described in this paper fare for these problems, and we predict varying degrees of success. We also stress that even utilising DSS will not allow the calculation of statistics at astrophysically realistic values. However it is to be hoped that, whereas the dynamics may be extremely sensitive to the parameters, for a range of problems the statistics may prove less so.  We believe 
that some of these problems may require the inclusion of higher order cumulants in the scheme and are currently engaged in determining efficient numerical procedures for their integration.

Clearly in the longer term, if these techniques prove useful for the simpler problems described above, it will be of interest to apply them to more computationally intensive problems in astrophysical fluids. These include the driving of zonal flows in planets, the mixing of angular momentum and abundances in stellar interiors and the transport by turbulence in accretion disks.

%\section{Figures}



% If in two-column mode, this environment will change to single-column
% format so that long equations can be displayed. Use
%\begin{widetext}
%\end{widetext}

% figures should be put into the text as floats.
% Use the graphics or graphicx packages (distributed with LaTeX2e)
% and the \includegraphics macro defined in those packages.
% See the LaTeX Graphics Companion by Michel Goosens, Sebastian Rahtz,
% entire page. There is no need to do explicit centering.
\begin{acknowledgments}
JBM thanks P. Kushner and T. Schneider for helpful discussions. 
The authors are grateful to F. Sabou and W. Strecker-Kellogg for assistance with the semi-implicit Krylov method.   
The work is supported in part by NSF grant DMR-0605619 (JBM and KD).  
\end{acknowledgments}

\end{document}